\documentclass[fleqn,usenatbib]{mnras}

\usepackage{newtxtext,newtxmath}

\usepackage[T1]{fontenc}

\DeclareRobustCommand{\VAN}[3]{#2}
\let\VANthebibliography\thebibliography
\def\thebibliography{\DeclareRobustCommand{\VAN}[3]{##3}\VANthebibliography}

\usepackage{graphicx}	
\usepackage{amsmath}	
\usepackage{subcaption}
\usepackage{physics}
\usepackage{float}
\usepackage[normalem]{ulem}

\newcommand{\bnew}{\textbf{B}}
\newcommand{\vnew}{\textbf{v}}

\newcommand{\bturbind}{\bnew^{\prime}}
\newcommand{\bmeanind}{\overline{\bnew}}
\newcommand{\vturbind}{\vnew^{\prime}}
\newcommand{\vmeanind}{\overline{\vnew}}

\newcommand{\mean}[1]{\overline{#1}}

\newcommand{\vect}[1]{\textbf{#1}}

\title[A new sub-grid model for MHD turbulence. I.]{Assessment of a new sub-grid model for magneto-hydrodynamical turbulence. I. Magnetorotational instability.}

\author[M. Miravet-Tenés, P. Cerd\'a-Dur\'an, M. Obergaulinger, and J.A.~Font]{
Miquel Miravet-Tenés,$^{1}$\thanks{E-mail: miquel.miravet@uv.es}
Pablo Cerdá-Durán,$^{1}$
Martin Obergaulinger$^{1}$, 
and José A. Font$^{1,2}$
\\
$^{1}$Departament d'Astronomia i Astrofísica, Universitat de València, C/ Dr Moliner 50, 46100, Burjassot (València), Spain\\
$^{2}$Observatori Astronòmic, Universitat de València, C/ Catedrático José Beltrán 2, 46980, Paterna (València), Spain
}

\date{Accepted XXX. Received YYY; in original form ZZZ}

\pubyear{2022}

\begin{document}
\label{firstpage}
\pagerange{\pageref{firstpage}--\pageref{lastpage}}
\maketitle

\begin{abstract}
Insufficient numerical resolution of grid-based, direct numerical simulations (DNS) hampers the development of instability-driven turbulence at small (unresolved) scales. As an alternative to DNS, sub-grid models can potentially reproduce the effects of turbulence at small scales in terms of the resolved scales, and hence can capture physical effects with less computational resources. We present a new sub-grid model, the {\it MHD-instability-induced-turbulence} (MInIT) mean-field model. MInIT is a physically motivated model based on the evolution of the turbulent (Maxwell, Reynolds, and Faraday) stress tensors and their relation with the turbulent energy densities of the magneto-rotational (MRI) and parasitic instabilities, modeled with two partial differential evolution equations with stiff source terms. Their solution allows obtaining the turbulent stress tensors through the constant coefficients that link them to the energy densities. The model is assessed using data from MRI in-box DNS and applying a filtering operation to compare the filtered data with that from the model. Using the $L_2$-norm as the metric for the comparison, we find less than one order-of-magnitude difference between the two sets of data. No dependence on filter size or length scale of unresolved scales is found, as opposed to results using the gradient model (which we also use to contrast our model) in which the $L_2$-norm of some of the stresses increases with filter size. We conclude that MInIT can help DNS by properly capturing small-scale turbulent stresses which has potential implications on the dynamics of highly-magnetized rotating compact objects, such as those formed during binary neutron star mergers.
\end{abstract}

\begin{keywords}
{(magnetohydrodynamics) MHD -- turbulence -- instabilities -- methods: numerical}
\end{keywords}



\section{Introduction}

Binary neutron star (BNS)
mergers have long been recognized as one of the most promising sources of gravitational radiation.
The detection of gravitational waves (GWs) from BNS mergers offers, among others, the possibility of constraining the equation of state (EOS) of dense matter at supranuclear densities. Two such mergers have so far been reported by the LIGO-Virgo-KAGRA collaboration, GW170817~\citep{GW170817} and GW190425~\citep{Abbott_2020}. The former was also famously observed by dozens of astronomical facilities across the electromagnetic (EM) spectrum, bringing into being the field of multi-messenger astrophysics~\citep{multimessenger_abbot}. The follow-up EM observations of GW170817 are consistent with the merging of two neutron stars that produce a black hole (BH) surrounded by an accretion disc, with an EM signature indicative of an $r$-process-induced optical transient known as a kilonova~\citep{Kasen:2017,Cowperthwaite:2017,Tanvir:2017}. 

BNS mergers have also long been regarded as progenitors of short gamma-ray bursts (sGRBs) \citep{1999ApJ...524..262M}, an association that has received significant support with the EM follow-up  observations of GW170817 \citep{multimessenger_abbot,2017ApJ...848L..13A}.
Before the final BH-torus system is realized, and depending on the total mass of the binary, the system may go through a short-lived phase in which a transient  post-merger object forms, a so-called hypermassive neutron star (HMNS). This object will eventually collapse to a BH once support against gravity by rotation or neutrino pressure lessens. It is believed that sGRBs stem from relativistic, magnetically-driven jets powered by the accretion disc of the BH.

Our understanding of the complex physics and dynamics involved in BNS mergers and in their post-merger evolution has significantly expanded in the last few decades. This has been possible thanks to the use of  numerical simulations, ever larger and more accurate in terms of computational resources and more sophisticated in terms of input physics. Numerical relativity is the tool which is employed to study these systems (see~\citet{Paschalidis:2017,Baiotti:2017,Duez:2019,Shibata:2019,Ciolfi:2020} for recent reviews). The huge parameter space of the problem, its dimensionality, and the amount of physics involved limit the number of simulations. Therefore, the description of the long-term evolution of the post-merger phase for generic initial conditions remains poorly constrained \citep{rezzolla,Shibata:2019,Ciolfi:2020}. As a result, linking the results of simulations with the data from multi-messenger observations of BNS mergers, sGRBs and kilonovae, is still to a large extent an ongoing task.

One key aspect of the riddle is the fate of the post-merger HMNS remnant whose lifetime can be strongly affected by a number of physical effects. Among the most significant ones is the turbulent amplification of the magnetic field occurring throughout the process, from the late inspiral to well inside the post-merger phase. While neutron stars in binary systems probably have a low magnetization ($B< 10^{12}$ G), magnetic fields can be largely amplified via instabilities such as the Kelvin-Helmholtz (KH) instability \citep{martin:2010,producing_magnetar_fields_giacomazzo,eff_mag_field_kiuchi_cerda,Kiuchi:2018} that develops in a shear layer when the neutron stars come into contact. This turbulence phase takes part during the first milliseconds of the merger and leads to a fast growth of small-scale magnetic fields \citep{Vigano:2020}. Another mechanism that acts on longer timescales is the magnetorotational instability (MRI) \citep{BALBUS_1991ApJ...376..214B,Balbus_Hawley__1998__RMP__MRI}. In the absence of further (de-)stabilizing effects such as gradients of entropy or composition, a weak magnetic field renders a fluid with a negative radial gradient of the angular velocity, $\Omega$, MRI-unstable. Small perturbations of the unstable layer grow exponentially on time scales given roughly by the rotational period. They take the form of channel modes, i.e., stacked layers in which the velocity and the magnetic fields have radial components of alternating polarity. The growth eventually terminates and the channels break down into small-scale turbulence. The precise conditions for the termination and thus the factor by which the seed magnetic field is amplified are not fully understood.

The presence of intense magnetic fields of ${\cal O}(10^{15})$G has important consequences in the dynamics of the post-merger remnant. The turbulence generated by MRI is probably the dominant process transporting angular momentum in the HMNS. The efficiency of this process is directly related with the timescale in which the BH forms and it leaves an imprint in the emitted GWs. Moreover, turbulent and convective dynamos could amplify the magnetic field and generate large-scale structures (dipole fields) that seem to favour the formation of jets and sGRBs. 

At merger the magnetic field is formed in extremely small spatial scales of ${\cal O}(10^{-2})$ m \cite{Guilet2017}.  This poses an enormous challenge for numerical simulations unable to resolve those scales with enough accuracy. Steps to overcome this limitation have been taken by \cite{2008PhRvD..78h4033B,2014CQGra..31g5012R,eff_mag_field_kiuchi_cerda,Kiuchi:2018,Kiuchi:2022,Murguia-Berthier:2021} through direct numerical magnetohydrodynamic (MHD) simulations\footnote{Although strictly speaking the term {\it direct numerical simulation} should be used exclusively for simulations in which all scales are resolved without the necessity of sub-grid models, here we use the term in a loose way to refer to  simulations without sub-grid models, even if not all scales are resolved.} with very high resolution.  Those revealed that a grid resolution of $\Delta \sim 10$ m is needed for the KH instability to efficiently amplify the magnetic field. In particular \cite{eff_mag_field_kiuchi_cerda} found that   KH-driven turbulence strongly amplifies the magnetic field energy by at least 6 orders of magnitude in  $\approx 4$ ms after merger. However, as these simulations are computationally very expensive (involving $\sim {\cal O}(10)$ million CPU hours) it is currently not possible to perform a systematic study of the magnetorotational evolution of BNS merger remnants. Moreover, it is not advisable to use low-resolution simulations since small-scale turbulence can feed the evolution of the magnetic field at large scales due to the emergence of inverse cascade phenomena. For this reason numerically unresolved simulations might lead to completely wrong results. 

An alternative way to direct numerical simulations is to use sub-grid models \citep{smagorinsky,Leonard:1975,mueller:2002,ogilvie,2017ApJ...838L...2R,2020Symm...12.1249R,2020MNRAS.491.5510W}. This allows performing relatively modest numerical simulations, in terms of resolution, by combining them with a model that describes the small-scale dynamics, smaller than the computational grid. Sub-grid models have found applications in astrophysics (e.g.~in stellar evolution \citep{book}) as well as in other fields of physics (e.g.~meteorology \citep{sub_grid_weather}) and engineering (e.g.~aerodynamics \citep{sub_grid_aerodyn}). The specific case of magnetized fluids comes with its own challenges, mostly related to the problem of the emergence of  inverse cascades. The majority of studies in this context have dealt with solar and stellar dynamos \citep{book}. In the last few years  there have been attempts to use sub-grid models in general relativity based on simple approaches~\citep{producing_magnetar_fields_giacomazzo,general_rel_kiuchi_shibata,Shibata:2021,extension_subgrid_vigano,paper_gradient_test}. Pioneering large-eddy simulations of BNS mergers have already been performed using some of those models~\citep{Vigano:2020,2021arXiv211208413P,Aguilera-Miret:2022}. 

In this paper we take first steps towards the development and testing of new approaches for sub-grid models for MHD and compare our proposal with current procedures based on the gradient sub-grid model~\citep{extension_subgrid_vigano,paper_gradient_test}. We restrict ourselves to the case of high conductivity and density, for which we have high Reynolds numbers and the fluid approximation is valid; this is precisely the case of interest for BNS mergers. The gradient model, based on the Taylor expansion and the inverse function theorem, is widely used nowadays due to the fact that it does not rely on any phenomenological assumption.
We first introduce sub-grid modelling with the $\alpha$-$\beta$ dynamo approach {from \citet{Parker1955} \citep[see][for details]{krause}} and perform an a priori test of the gradient model by computing the Pearson correlation coefficient to check for linearity between the model and the data from an in-box  MRI simulation. The same numerical data is used to assess our new  proposal. The sub-grid model we put forward in this work is based on the proportionality relations between the components of the turbulent stress tensors. We devise evolution equations for the turbulent energy densities \citep{paper_simulation,Pessah_2009,Pessah_2010} that make it straightforward to model the stress tensors in terms of the energy densities of the MRI and the parasitic instabilities. The form of the evolution equation depends  on the physics of the particular instability under consideration, which, in the case of this work will be the MRI. A performance comparison between our model and the gradient sub-grid model of~\cite{paper_gradient_test} is done by computing the $L_2$ relative error norm for different filter sizes and grid resolutions. 

This paper is organized as follows: in Section~\ref{sect: sec2} we discuss the mean-field formalism used to separate numerically resolved quantities from the small-scale turbulent ones. Next, in Section~\ref{sect: sec3} we show a direct application of the mean-field formalism to the induction equation with the determination of the $\alpha$ and $\beta$ dynamo coefficients, and present the basis of both, the gradient sub-grid model and our new model. In Section~\ref{sect: sec4} we briefly describe the MRI simulation we use to carry out our testing of the new sub-grid method and the model comparison. The results of the various tests are also reported in this section. Finally, our conclusions are summarized in Section~\ref{sect: sec5}. Some equations in the manuscript contain indices with Latin characters. Those are  spatial and hence take values from $1$ to $3$. Unless stated otherwise we use geometrized units by setting $G = c = 1$, and the magnetic permeability is also set to $\mu_0=1$, corresponding to a Gaussian or Heaviside-Lorentz unit system \citep[see][]{Jackson1975}.

\section{Mean-field MHD}
\label{sect: sec2}

\subsection{Newtonian MHD equations}

The mathematical framework for our study is Newtonian MHD whose equations we review in this section. The MHD equations stem from the result of applying the Navier-Stokes and the Maxwell equations to an electrically conducting fluid (or plasma). The equations couple the different dynamical variables of the plasma, such as the fluid velocity, the gas pressure, the mass density and the magnetic field. A Newtonian approach can be applied when the plasma velocity is not relativistic. A common further simplification is to consider the ideal MHD case where the fluid has an infinite electric conductivity, $\sigma \rightarrow \infty$. In this case Ohm's law reduces to
\begin{equation}\label{ohms_law}
 E_i = -\epsilon_{ijk}v_j B_k\,,
\end{equation}
where $\epsilon_{ijk}$ is the 3-dimensional Levi-Civita symbol. This means that the electric field \textbf{E} is completely determined by the magnetic field \textbf{B} and the fluid velocity \textbf{v}. Therefore, by 
inserting Eq.~\eqref{ohms_law} into 
Faraday's law
\begin{equation}
    \curl \textbf{E} = -\frac{\partial \textbf{B}}{\partial t},
\end{equation}
one obtains the induction equation:
\begin{equation}\label{induction_vect}
 \frac{\partial \textbf{B}}{\partial t} = \curl(\textbf{v}\times \textbf{B}). 
\end{equation}
If we use the expansion of the curl of the vector product
\begin{equation}
    \curl (\textbf{v}\times\textbf{B}) = \textbf{v}(\div\textbf{B})-\textbf{B}(\div\textbf{v})+(\textbf{B}\cdot \gradient)\textbf{v}-(\textbf{v}\cdot \gradient)\textbf{B},
\end{equation}
along with the solenoidal condition of the magnetic field, $\div \textbf{B} = 0$, one can rewrite the induction equation as
\begin{equation}\label{induction_eq}
  \partial_t B^i + \partial_j (v^jB^i-v^iB^j) = 0\,.  
\end{equation}
This is the first equation of the MHD system. The remaining equations follow immediately from the equations of 
mass continuity, the Euler equation, and the energy equation. The final set of MHD equations can be cast in the following conservation form:
\begin{equation}\label{mhd_eqs}
   \partial_t\textbf{F}^0+\partial_j\textbf{F}^j = 0\,, 
\end{equation}
where vector $\textbf{F}^0 \equiv \textbf{C}$ is the state vector, whose components are the following conserved quantities 
\begin{equation}\label{state_vect}
  \textbf{C} =   \left[\begin{array}{c}
         \rho  \\
        N^i \\
        U \\
        B^i 
        \end{array} \right],
\end{equation}
which correspond to the mass density, the momentum density, the energy density, and the magnetic field, respectively. The vectors $\textbf{F}^j$ are the fluxes along spatial direction $j$, 
\begin{equation}\label{flux_vect}
  \textbf{F}^j \equiv \left[\begin{array}{c}
    N^j \\
    T^{ji} \\
    S^j \\
    D^{ji}   
  \end{array} \right] =   \left[\begin{array}{c}
         \rho v^j \\
         \rho v^iv^j-B^iB^j+\delta^{ij}\big[p+B^2/2\big] \\
        v^j\big[U+p+B^2/2\big]-(v_kB^k)B^j \\
        v^jB^i-v^iB^j 
        \end{array} \right],
\end{equation}
where $p$ is the thermal pressure and $B^2=B^iB_i$.

The fluxes are written in terms of the \textit{primitive quantities}, $ P = \lbrace \rho,v^i,\varepsilon, B^i\rbrace$, where 
$\varepsilon$ is the specific internal energy. One can express the conserved fields $\textbf{C}$ in terms of the primitive ones\footnote{Note that the mass density and the magnetic field are both primitive and conserved fields.}: 
\begin{subequations}
\begin{align}
N^i & = \rho v^i , \\
U & = \rho(\varepsilon+v^2/2)+B^2/2\,.
\end{align}
\end{subequations}

\subsection{Foundations of mean-field MHD}

The aim of this work is to employ the mean-field MHD formalism to the previous equations. In this approach a filter is applied to all variables both in space and time, over certain characteristic (small) length to compute mean quantities. Any variable can then be decomposed into a mean and a fluctuating (turbulent) component of zero mean. Thus, the MHD equations can be written in terms of mean quantities (resolved scales in numerical simulations) and of the average of combinations of the fluctuations (unresolved scales). A sub-grid model provides a closure relation between the two terms that allows to write the system of equations as a closed system amenable to be solved numerically. 

Given any field $\textbf{A}$, the corresponding mean field will be denoted by  $\overline{\textbf{A}}$, defined to be the expectation value of $\textbf{A}$ in an ensemble of identical systems. The averages that will be used in the MHD equations can be both spatial or temporal. Hence, the averaging operator will be defined following \cite{book} as either 
\begin{equation}
    \overline{\textbf{A}} = \frac{1}{V}\int_{V} \textbf{A}(t,\textbf{x})\, d^3\textbf{x}\,,
\end{equation}
for a spatial average in a scale of order $\lambda$ (having thus a volume $V \propto \lambda^3$), or
\begin{equation}
	\overline{\textbf{A}} = \frac{1}{\tau}\int_{\tau} \textbf{A}(t,\textbf{x}) \, dt \,,
\end{equation}
for a time average in a timescale $\tau$. Moreover, the average can be performed over different realizations of a simulation. We denote by $\textbf{A}^{\prime}$ the difference between the original field and the mean field,
\begin{equation}
   \textbf{A}^{\prime} = \textbf{A}-\overline{\textbf{A}}\,.
\end{equation}
We will refer to it as the {\it fluctuating field}. 
The above decomposition can be physically interpreted as follows: the field $\textbf{A}$ is characterized by a slowly varying component, $\overline{\textbf{A}}$, which varies on a  large spatial (temporal) scale $L$ ($T$) and is properly resolved, plus a rapidly fluctuating part, $\textbf{A}^{\prime}$, which varies on a much smaller (shorter) scale, \textit{l} (\textit{t}), and represents the effect of the unresolved dynamics. Therefore, the average operation is computed over an intermediate spatial (temporal) scale $\lambda \, (\tau)$: $\textit{l} \ll \lambda \ll L$ ($\textit{t} \ll \tau \ll T$). 

 It is useful to consider the following relations, known as the Reynolds averaging relations~\citep{krause},
\begin{equation}\label{reynolds_relations}
\begin{split}
    \textbf{A} = \overline{\textbf{A}}+\textbf{A}^{\prime}, \hspace{0.5cm} \overline{\overline{\textbf{A}}} & = \overline{\textbf{A}}, \hspace{0.5cm} \overline{\textbf{A}^{\prime}} = 0, \\
    \overline{\textbf{A}+\textbf{C}} =\overline{\textbf{A}}+\overline{\textbf{C}}, \hspace{0.5cm} \overline{\overline{\textbf{A}}\overline{\textbf{C}}} & = \overline{\textbf{A}}\overline{\textbf{C}}, \hspace{0.5cm} \overline{\overline{\textbf{A}}\textbf{C}^{\prime}} = 0, 
\end{split}
\end{equation}
holding for any given fields $\textbf{A}$ and $\textbf{C}$. 

Since the average of the fluctuating part is zero, the only possible terms related to the fluctuating part in the mean-field equations are the mean of combinations of two fluctuating variables. For this work, we consider only fluctuations of the velocity ($\vturbind$) and of the magnetic field ($\bturbind$), and neglect the fluctuations of density ($\rho^\prime=0$) and internal energy ($\epsilon^\prime=0$). This is similar to the work of \citet{ogilvie} that consider the incompressible shearing sheet case \citep{GOLDREICH_1965MNRAS}. {The incompressible case is also a very good approximation for the particular case of the MRI (see \cite{GOODMAN_1994ApJ...432..213G} and the discussion in \cite{paper_simulation})} With these considerations, the fluctuating part in the equations can be written in terms of the following tensors:
\begin{subequations}\label{stress_tensors}
\begin{align}
	M_{ij} & = B^{\prime}_i B^{\prime}_j,\\
	R_{ij} & = v^{\prime}_i v^{\prime}_j, \\
	F_{ij} & = v^{\prime}_i B^{\prime}_j-v^{\prime}_j B^{\prime}_i\,, 
\end{align}
\end{subequations}
namely, the Maxwell, Reynolds and Faraday stress tensor, respectively, as defined in \cite{ogilvie}. From their definitions, it follows that the Maxwell and Reynolds tensors are symmetric, while the Faraday stress tensor is antisymmetric. These objects naturally appear when performing the averaging of the MHD equations, which have the form \citep{ogilvie}
\begin{equation}\label{mf_mhd_eqs}
   \partial_t\widetilde{\textbf{F}}^0+\partial_j\widetilde{\textbf{F}}^j = \textbf{S}(\overline{R_{ij}},\overline{M_{ij}},\overline{F_{ij}})\,, 
\end{equation}
where quantities with tilde refer to identical functional expressions as Eqs.~\eqref{state_vect} and \eqref{flux_vect}, but for the mean quantities $\overline{\textbf{B}}$ and $\overline{\textbf{v}}$ instead of $\textbf{B}$ and $\textbf{v}$. Note that $\widetilde{\textbf{F}}^0 \ne \overline{\textbf{F}}^0$ and $\widetilde{\textbf{F}}^j \ne \overline{\textbf{F}}^j$. The source term $\mathbf{S}$ is a function of the mean stresses. Its value for some particular cases can be found e.g. in \cite{krause} or \cite{ogilvie}. The general case is not of direct interest for this work and will be presented elsewhere. In Eq.~(\ref{mf_mhd_eqs}) only quadratic terms, $\mathcal{O}(\rm{\textbf{A}}^{\prime 2})$, have been considered, and higher order terms have been neglected. This should be a good approximation if $|\textbf{A}^\prime|^2\ll|\overline{\textbf{A}}|^2$.

\section{Sub-grid models}
\label{sect: sec3}

We now turn to discuss some sub-grid models that have already been used in previous works, along with the new model we propose in this paper.

\subsection{$\alpha$-$\beta$ dynamo mean-field model}
\label{sect:dynamos}

Let us start by filtering the induction equation given in Eq.~\eqref{induction_vect}. First, we express the fields in terms of resolved and unresolved parts: 
\begin{equation}\label{mean_turb_ind_eq}
	\dfrac{\partial}{\partial t}(\bmeanind+\bturbind) = \curl [(\vmeanind+\vturbind)\times(\bmeanind+\bturbind)]\,.  
\end{equation}
Averaging the previous expression leads to: 
\begin{equation}\label{averaged_induction_eq}
	\dfrac{\partial}{\partial t}\bmeanind = \curl (\vmeanind \times \bmeanind)+\curl(\overline{\vturbind\times\bturbind})\,, 
\end{equation}
where we have made use of $\overline{\bturbind}= \overline{\vturbind} = 0$ and $\overline{\vmeanind\times \bturbind} = \overline{\vturbind\times \bmeanind} = 0$. Note that Eq.~\eqref{averaged_induction_eq} is the same as Eq.~\eqref{induction_vect} but written in terms of filtered quantities and with an extra term. Also note that the contribution $\overline{\vturbind\times\bturbind}$ comes from non-linearity. Therefore, we define the turbulent electromotive force as: 
\begin{equation}\label{turb_emf}
 \xi \equiv \overline{\vturbind \times \bturbind} \,.
\end{equation}

The $\alpha$-$\beta$ dynamo mean-field model provides a closure equation for $\xi$ expressed in terms of the mean quantities. It assumes statistical homogeneity, steadiness and isotropy  for $\vturbind$. Due to the homogeneity of $\vturbind$, the electromotive force will only change with position as far as the mean magnetic flux density and the spatial derivatives do. Furthermore, due to the isotropy of $\vturbind$, any quantity derived from it must be rotation-invariant, and the only vector that does so is the zero-vector. Therefore, there is no contribution of $\vturbind$ to the vector structure of $\xi$. Additionally, it assumes that $\bmeanind$ depends so weakly on time and position that $\xi$ can be represented by $\bmeanind$ and its first spatial derivatives. Under these assumptions, the turbulent electromotive force has to fulfill
\begin{equation}\label{turb_force_expansion}
  \xi_i = \alpha_{\textrm{d }\, ij} \overline{B}^j-\beta_{\textrm{d }\, ij}(\curl \overline{B})^j, 
\end{equation}
where $\alpha_{\rm{d}}$ and $\beta_{\rm{d}}$ are the so-called dynamo coefficients that depend on the turbulent velocity field, $\vturbind$ \citep{book,reboul-salze_mri}. 
This equation can be used as a closure relation that allows us to express Eq.~\eqref{averaged_induction_eq} solely in terms of resolved quantities ($\bmeanind$, $\vmeanind$).

In the numerical simulations of MRI of \cite{reboul-salze_mri} no correlation was found between the electromotive force from Eq.~\eqref{turb_force_expansion} and the mean current $\overline{J} = -\curl\overline{B}$. Therefore, one could simplify equation~\eqref{turb_force_expansion} to
\begin{equation}\label{final_close_alpha}
   \xi_i = \alpha_{\textrm{d }\, ij}\overline{B}^j\,. 
\end{equation}
This closure relation is one of the sub-grid models we test in this work. The interested reader is addressed to \cite{krause} and \cite{reboul-salze_mri} for further information.

\subsection{Gradient sub-grid model}

The gradient sub-grid  model~\citep{Leonard:1975,mueller:2002} has  recently received attention in compressible MHD studies, in particular to investigate magnetic-field amplification in BNS mergers~\citep{extension_subgrid_vigano,paper_gradient_test,Vigano:2020,Aguilera-Miret:2020,2021arXiv211208413P,Aguilera-Miret:2022}. In this model one does not need to assume a phenomenological form for the sub-filter-scale terms as it simply relies on the Taylor expansion of the fluxes of the MHD equations, Eq.~\eqref{flux_vect}, expressed in terms of the primitive variables.

We start by writing the primitive fields $P$ in terms of the conserved ones, Eq.~\eqref{state_vect}, by computing $\widetilde{P}^l \equiv g^l(\overline{\textbf{C}})$. Here, the "$\sim$" symbol over a given field means that this field is expressed in terms of filtered quantities. In our case, the primitive variables will be thus functions of the filtered conserved fields. Therefore, we obtain the following expression when applying a filter over the MHD equations: 
\begin{equation}\label{filtered_eq_ex}
    \partial_t \overline{C}^{l} + \partial_j \overline{F}^{j,l} = 0\,,
\end{equation}
where index "\textit{j}" represents the spatial directions and index "\textit{l}" represents the fields of each set of quantities. We next express the fluxes in terms of $\widetilde{P}\big(\overline{\textbf{C}}\big)$, which leads to: 
\begin{equation}\label{filtered_eq_with_sfs}
     \partial_t \overline{C}^{l} + \partial_j F^{j,l}(\widetilde{P}) = \partial_j {\tau}_{\rm F}^{j,l}\, . 
\end{equation}
The term on the right-hand side is the sub-filter scale (SFS) tensor, \begin{equation}\label{sfs_tensors}
   {\tau}_{\rm F}^{j,l} \equiv F^{j,l}(\widetilde{P})-\overline{F^{j,l}(P)} \, ,
\end{equation}
and there will be one for each flux. In order to express these tensors only in terms of the filtered variables and their derivatives, one performs a Taylor expansion to first order in $\eta$ (which is related to the filter size) of $\overline{\textbf{F}^{j,l}(P)}$ around $\mean{P}$
\begin{equation}
    \overline{F^{j,l}(P)} \simeq F^{j,l}(\mean{P}) + \eta\left(\nabla^2 F^{j,l}(\mean{P})-\frac{dF^{j,l}}{d\mean{P}^m}\nabla ^2\mean{P}^m\right) \,,
\end{equation}
and then expand around $\widetilde{P}$
\begin{equation}
    F^{j,l}(\mean{P}) \simeq F^{j,l}(\widetilde{P}) +\eta \frac{dF^{j,l}}{d\widetilde{P}^m}\left(\nabla^2 \widetilde{P}^m-\frac{d\widetilde{P}^m}{d\mean{C}^n}\nabla^2\mean{C}^n\right)\,,
\end{equation}
where the indices $(l,m,n)$ denote the spatial components of fields. By the inverse function theorem we are able to express the primitive variables, $P$, in terms of the conserved ones, $\widetilde{P}=P(\mean{C})$ (the full procedure is outlined in \cite{paper_gradient_test}). Finally, we can re-express the SFS tensors from Eq.~\eqref{sfs_tensors} in the form 
\begin{equation}\label{gradient_sfs_tensors}
    \tau_{\rm F}^{j,l} = -\eta\nabla \frac{d F^{j,l}}{d\mean{C}^n}\cdot\nabla \mean{C}^n\,.
\end{equation}

This is the closure relation of the gradient sub-grid model. It relates the contribution of the dynamics from the smaller scales to the filtered variables and their derivatives. The SFS tensor in Eq.~\eqref{gradient_sfs_tensors} is proportional to the filter size, i.e. the typical scale of the simulation, given by the numerical resolution. The accuracy of the derivatives of the filtered variables depends on the numerical method used to compute them and also on the resolution of the simulation. Therefore, the model is expected to perform better for higher resolutions and also for smaller sizes of the filter \citep{paper_gradient_test}. 

The filtered version of the MHD system in terms of the conserved variables $\mean{C}^m$ reads:
\begin{subequations}\label{filtered_mhd_system}
\begin{align}
   \partial_t & \mean{\rho} +\partial_j {N}^j(\widetilde{P}) =  \partial_j\tau^j_{\rm N} , \label{filt_continuity} \\
   \partial_t & \mean{N}^i+\partial_j T^{ji}(\widetilde{P}) = \partial_j \tau^{ji}_{\rm T}, \label{filt_euler} \\
   \partial_t & \mean{U} +\partial_j S^j(\widetilde{P}) = \partial_j\tau^j_{\rm S}, \label{filt_energy} \\
   \partial_t & \mean{B}^i+\partial_j D^{ji}(\widetilde{P}) = \partial_j\tau^{ji}_{\rm D} \label{filt_induction}\,.
\end{align}
\end{subequations}
These equations have the same form as Eq.~\eqref{filtered_eq_with_sfs} and the non-negligible contribution of the smaller scales appears in the form of source terms in the rhs of the equations. Thanks to the closure relation for the SFS tensors given by Eq.~\eqref{gradient_sfs_tensors}, the source terms can be computed with the derivatives of the fluxes with respect to the conserved variables. Their final expressions are:
\begin{subequations}\label{sfs_tensors_mhd}
\begin{align}
    \tau_{\rm N}^j & = - \eta \nabla \frac{d\mean{N}^j}{d\mean{C}^m}\cdot \nabla \mean{C}^m = 0 \,, \label{tau_N} \\
    \tau^{ji}_{\rm T} & = - \eta \nabla \frac{dT^{ji}}{d\mean{C}^m}\cdot \nabla \mean{C}^m = \eta \Big[-2 \mean{\rho} \nabla \widetilde{v}^i\cdot \nabla\widetilde{v}^j+2\nabla \widetilde{B}^i\cdot \nabla\widetilde{B}^j- \nonumber \\
    - & \delta^{ji} \big[\nabla\frac{d\widetilde{P}}{d\mean{\rho}}\cdot\nabla\mean{\rho}+\nabla\frac{d\widetilde{P}}{d\widetilde{\varepsilon}}\cdot\nabla\widetilde{\varepsilon}-\frac{2}{\mean{\rho}}\frac{d\widetilde{P}}{d\widetilde{\varepsilon}}\nabla\mean{\rho}\cdot\nabla\widetilde{\varepsilon}+\nabla\mean{B}_k\cdot\nabla\mean{B}^k- \nonumber \\
    - & \frac{1}{\mean{\rho}}  \frac{d\widetilde{P}}{d\widetilde{\varepsilon}}\big(\mean{\rho}\nabla\widetilde{v}_k\cdot\widetilde{v}^k+\nabla\mean{B}_k\cdot\nabla\mean{B}^k\big)\big] \Big] \,, \label{tau_T} \\
  \tau^j_{\rm S} & = - \eta \nabla \frac{d\mean{S}^j}{d\mean{C}^m}\cdot \nabla \mean{C}^m = \eta \Big[ -2\big[\nabla\widetilde{\Theta}\cdot\nabla\widetilde{v}^j+\big(\mean{B}^j\mean{B}_k\nabla\widetilde{v}^k- \nonumber \\
   - & \widetilde{\Theta}  \nabla\widetilde{v}^j\big)\cdot \nabla(\ln\mean{\rho})-\mean{B}^j\nabla\mean{B}_k\cdot\nabla\widetilde{v}^k-\nabla(\widetilde{\textbf{v}}\cdot\mean{\textbf{B}})\cdot \nabla\mean{B}^j\big]+\nonumber \\
  +&\widetilde{v}^j\big[\big[\nabla\frac{d\widetilde{P}}{d\mean{\rho}}\cdot\nabla\mean{\rho}+\nabla\frac{d\widetilde{P}}{d\widetilde{\varepsilon}}\cdot\nabla\widetilde{\varepsilon}-\frac{2}{\mean{\rho}}\frac{d\widetilde{P}}{d\widetilde{\varepsilon}}\nabla\mean{\rho}\cdot\nabla\widetilde{\varepsilon}+\nabla\mean{B}_k\cdot\nabla\mean{B}^k- \nonumber \\
   - & \frac{1}{\mean{\rho}} \frac{d\widetilde{P}}{\widetilde{\varepsilon}}\big(\mean{\rho}\nabla\widetilde{v}_k\cdot\widetilde{v}^k+\nabla\mean{B}_k\cdot\nabla\mean{B}^k\big)\big]\Big] \,, \label{tau_S} \\
 \tau^{ji}_{\rm D} & = - \eta \nabla \frac{dD^{ji}}{d\mean{C}^m}\cdot \nabla \mean{C}^m = -2\eta\big[\nabla\widetilde{v}^i\cdot\nabla\mean{B}^j-\nabla\widetilde{v}^j\cdot\nabla\mean{B}^i+ \nonumber \\
 + & \big(\mean{B}^i\nabla\widetilde{v}^j-\mean{B}^j\nabla\widetilde{v}^i\big)\cdot\nabla(\ln \mean{\rho})\big] \,, \label{tau_D} 
\end{align}
\end{subequations}
where $\widetilde{\Theta} \equiv \mean{U}+\widetilde{p}+\mean{\textbf{B}}^2/2$. The derivatives of $\widetilde{p}$ with respect to the mass density and the specific internal energy can be obtained using the equation of state. 

As we show below in Sec.~\ref{sect: sec4}, to assess the method we will employ the SFS tensors given by Eqs.~\eqref{sfs_tensors_mhd} to results of in-box MRI numerical simulations, comparing their values with those computed using Eq.~\eqref{sfs_tensors}.

\subsection{MHD-instability-induced-turbulence (MInIT) mean-field model)}

In the previous sections we have shown that the sub-grid-scale terms that arise from averaging the MHD equations can be modeled in terms of averaged variables. For the gradient sub-grid model, the SFS terms represent the sub-grid contributions of the fluxes (cf. Eq.~\eqref{sfs_tensors}). In fact, those terms have been obtained by taking gradients of the derivatives of the fluxes with respect to the conserved variables (see Eq.~\eqref{gradient_sfs_tensors}). 

In our {new} model the goal is similar, i.e.~we want to express the contributions of the sub-grid scales in terms of resolved quantities that evolve over large enough scales so that their evolution can be captured by direct numerical simulations. \cite{ogilvie} proposed a mean-field model based in the computation of the evolution equations for the mean stresses of the form
\begin{align}
    (\partial_t + \overline{v}_k \partial_k ) \overline{M}_{ij} - \overline{M}_{ik} \partial_k \overline{v}_j  - \overline{M}_{jk} \partial_k \overline{v}_i &= S^{(M)}_{ij}, \label{eq:ogilvieM}\\
    (\partial_t + \overline{v}_k \partial_k ) \overline{R}_{ij} - \overline{R}_{ik} \partial_k \overline{v}_j  - \overline{R}_{jk} \partial_k \overline{v}_i &= S^{(R)}_{ij}, \label{eq:ogilvieR} 
\end{align}
where $S^{(M)}_{ij}$ and $S^{(R)}_{ij}$ are functions depending on the averages of combinations of three fluctuating variables (e.g. $\overline{v_i v_j b_k}$ ... ), and could be approximated by a closure relation. The left-hand-side consists of an advective term and a term accounting for the ``stretching'' by gradients of the mean velocity. \cite{ogilvie} considered only the particular case of $\overline{F}_{ij}=0$ but, in the most general case, this quantity should fulfill analogous equations.

One could try to find a general closure for the system of equations proposed by \cite{ogilvie} (or a generalization of this system) but this approach is in general complicated, and the number of additional equations to be solved is significant ($12$ equations just for the independent components in Eqs.~\eqref{eq:ogilvieM} and \eqref{eq:ogilvieR}).
Instead of this general approach, we aim at providing a sub-grid model that resolves turbulence induced by MHD instabilities. The most relevant MHD instabilities developed during BNS mergers are the magneto-rotational instability and the Kelvin-Helmholtz instability. In this work we focus on modelling the former deferring to a following study the treatment of the latter. 

The MRI has been studied through numerical simulations by a number of authors (see references in the Introduction). Here we consider the semi-global simulations of \cite{Rembiasz-2016} and use those as part of our tests and model calibration in the next sections. The simulations are discussed in detail in Section~\ref{simulations} below. Fig.~\ref{fig:stresses} displays the time evolution of the stress tensor components in one such simulation. All three tensors evolve in a qualitatively similar way. On top of the initial conditions (a differentially rotating fluid with a vertical magnetic field) the instability grows exponentially developing non-linear channel flows. Those are eventually disrupted leading to the termination of the exponential growth. After termination, the fluid settles into a turbulent state in which stresses are approximately constant (in a statistical sense). This behaviour can be understood by the parasitic instability (PI) theory \citep{GOODMAN_1994ApJ...432..213G,Pessah_2009}. In this model, the channel flows, which are exact solutions of the non-linear incompressible MHD equations \citep{GOODMAN_1994ApJ...432..213G,Pessah_2008}, are disrupted by secondary (parasitic) instabilities growing in time. In the case of high Reynolds numbers those PI are of Kelvin-Helmholtz type, and their growth rate depends on the exponentially growing shear, leading to a super-exponential growth. As the energy of the PI becomes comparable to that of the channel flows (MRI energy), those are disrupted and the balance between the two instabilities settles the system into a turbulent state. \cite{Rembiasz-2016} were able to measure the PI energy and its super-exponential growth, and tested the validity of the termination criterion giving strong support to the PI theory.

\begin{figure}
    \centering
    \includegraphics[width=0.9\linewidth]{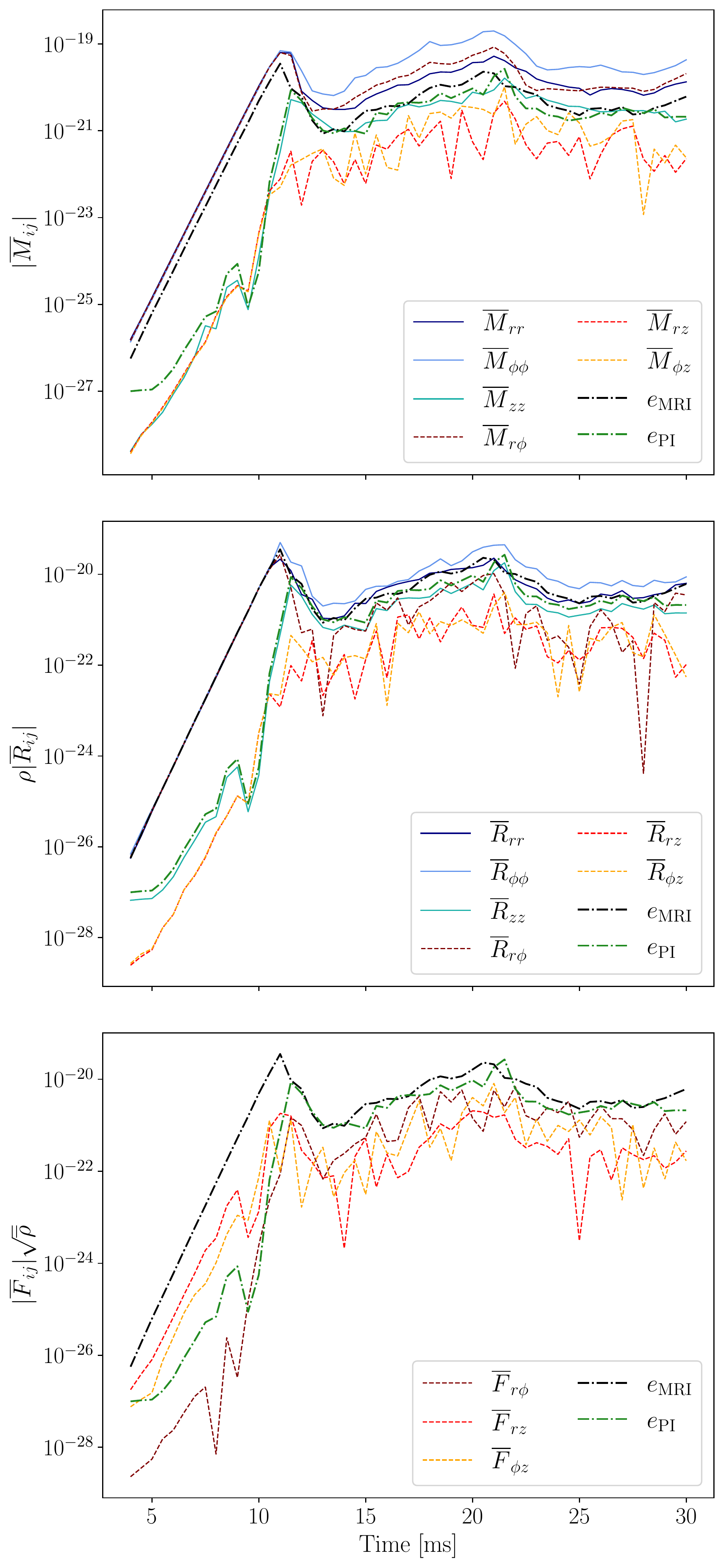}
    \caption{Evolution of the Maxwell (top), Reynolds (middle) and Faraday (bottom) stress tensors computed from the numerical data of the simulation MRI-H1 performed by \citet{Rembiasz-2016}. The dash-dotted black and green lines represent the energy densities of the stresses associated with the MRI and PI, respectively.}
    \label{fig:stresses}
\end{figure}

Fig.~\ref{fig:stresses} compares the different stresses for a particular numerical simulation with an estimation of the energy densities of the MRI, $e_{\rm MRI}$, and of the PI, $e_{\rm PI}$ (see next section for definitions). The general trend is that, up to some constants, $e_{\rm MRI}$ and $e_{\rm PI}$ are good tracers of the evolution of all stresses. Similar behaviour is observed in all the simulations analyzed. This motivates the development of a mean-field model based upon evolution equations for $e_{\rm MRI}$ and $e_{\rm PI}$ so that the stresses can be computed from these quantities using calibrated constants (closure relation). In the next four subsections we discuss the MRI and PI and provide the details of the two ingredients (closure relations and energy density equations) that conform the {\it MDH-instability-induced-turbulence (MInIT) mean-field model}, the main subject of this work.

\subsubsection{MRI theory}

We consider a rotating fluid with a magnetic field with non-zero vertical component\footnote{By ``vertical'' it must be understood the component of the magnetic field parallel to the rotational axis, i.e.~along the $z$-direction of the cylindrical coordinate system $(r, \phi, z)$.}. We also consider the case of high Reynolds and magnetic Reynolds numbers (effectively ideal MHD for MRI fluctuations). In this case, the general dispersion relation for the modes of wave-like perturbations is \citep{Balbus1995,OBERGAULINGER_2009A&A...498..241O}: 
\begin{equation}\label{general_disp_rel}
    (\hat{\gamma}^2-\hat{k}^2)^2-(\hat{\gamma}^2-\hat{k}^2)(\hat{\omega}_{\rm G}^2+\hat{\omega}_{\rm R}^2 +4\cos^2{\theta_k})-4\hat{k}^2\cos^2{\theta_k} = 0 \,,
\end{equation}
where $\theta_k$ is the angle between the wavevector $\vect{k}$ and the vertical axis, $\hat{\omega}_{\rm G, R}$ are dimensionless frequencies related to buoyancy terms and differential rotation, respectively, and $\hat{\gamma} = \gamma/\Omega$ is the dimensionless growth rate of the mode, normalized by the rotational profile
\begin{equation}\label{ang_velocity}
    \Omega = \Omega_0\Big(\frac{r}{r_0}\Big)^{-q}\,,
\end{equation}
with $\Omega_0$ being the angular velocity at the characteristic radius $r_0$ and 
\begin{equation}
    q=-\frac{\rm{d}\ln{\Omega}}{\rm{d}\ln{r}} \,, 
\end{equation}
corresponding to the rotational shear. In Eq. \eqref{general_disp_rel} we introduce $\hat{k} = \vect{k}\cdot \vect{c}_{\rm A}/\Omega$, where $\vect{c}_{\rm A}$ is the Alfvén speed. 

Now, let us focus on the case without buoyancy effects ($\hat{\omega}_{\rm G}=0$). In this case, the fastest growing mode has a vertical wavevector \citep{paper_simulation}:
\begin{equation}\label{mri_wavevect}
      k_{\rm MRI} \equiv k_{\rm crit} = \sqrt{1-\frac{(2-q)^2}{4}}\frac{\Omega}{c_{\rm{A}z}}\,,
\end{equation}
where $c_{\rm{A}z}=\overline{B}_{z}/\sqrt{\rho}$ is the vertical component of the Alfvén velocity and is given by the initial magnetic field amplitude in the vertical direction. Thus, the expression of $\hat{k}$ is simplified for the fastest growing mode to
\begin{equation}
    \hat{k}_{\rm MRI} = \sqrt{1-\frac{(2-q)^2}{4}}\,.
\end{equation}
For the fastest growing mode the perturbations lie on the $r\phi$ plane and therefore $\theta_{k, {\rm crit}} = 0$. This implies that the dispersion relation given by Eq.~\eqref{general_disp_rel} leads to the following growth rate of the MRI: 
\begin{equation}\label{growth_rate_mri}
    \gamma_{\rm MRI} = \frac{q}{2}\Omega\,,
\end{equation}
which is constant in time. 

\cite{GOODMAN_1994ApJ...432..213G} demonstrated that this kind of perturbative solutions are not only solution of the linearized MHD equations but also of the full non-linear equations in the incompressible limit, giving rise to channel flows. During this phase of exponential growth  channel modes are characterized very accurately by its perturbation velocity and magnetic field, which we identify as $\mathbf{v}^\prime$ and $\mathbf{B}^\prime$, fulfilling the next properties in the ideal MHD case \citep[c.f.][]{Pessah_2008}
\begin{align}
    \mathbf{v}^\prime &=  v_0 \left ( \hat{\mathbf{r}} + \hat{\mathbf{\phi}} \right ) \, \sin{(k z)} \, e^{\gamma t} \,, \\
    \mathbf{B}^\prime &=  B_0 \left ( \hat{\mathbf{r}} - \hat{\mathbf{\phi}} \right ) \, \cos{(k z)} \, e^{\gamma t} \,,
\end{align}
where $\gamma$ and $k$ refer to the fastest growing mode and $v_0$ and $B_0$ are the velocity and magnetic field channel-flow amplitudes which are related by
\begin{equation}
    \frac{B_0}{\sqrt{\rho}} = \sqrt{\frac{4-q}{q}} \, v_0\,.
\end{equation}
Note that $\overline{{\mathbf{B}^\prime}^2}= B_0^2\, e^{\gamma t}$ and $\overline{{\mathbf{v}^\prime}^2}=v_0^2 \, e^{\gamma t}$. These expressions allow to write the contribution of the MRI channel flows to all the stresses in terms of $v_r^\prime$:
\begin{align}
&    \overline{R}^{\rm MRI}_{rr}=\overline{R}^{\rm MRI}_{\phi\phi}=\overline{R}^{\rm MRI}_{r\phi} &=& \,\,\,\,\overline{v_r^\prime v_r^\prime}\,,&\\
&    \overline{R}^{\rm MRI}_{rz}=\overline{R}^{\rm MRI}_{\phi z}=\overline{R}^{\rm MRI}_{zz} &=& \,\,\,\,0\,,&\\
&    \overline{M}^{\rm MRI}_{rr}=\overline{M}^{\rm MRI}_{\phi\phi}=-\overline{M}^{\rm MRI}_{r\phi} &=& \,\,\,\,(4/q-1)\,\rho\,\, \overline{v_r^\prime v_r^\prime}\,,&\\
&    \overline{M}^{\rm MRI}_{rz}=\overline{M}^{\rm MRI}_{\phi z}=\overline{M}^{\rm MRI}_{zz} &=& \,\,\,\,0\,,&\\
&    \overline{F}^{\rm MRI}_{ij}&=& \,\,\,\,0\,,& 
\end{align}
where we have used that $\overline{\sin^2{(k z)}}=\overline{\cos^2{(k z)}}=1/2$ and $\overline{\sin{(k z)}\cos{(k z)}}=0$ when averaging over the appropriate lengthscale (larger than $\lambda_{\rm MRI}$).
The contribution of the MRI channel flows to the (kinetic) energy density is then just
\begin{equation}
    e_{\rm MRI} = \frac{1}{2} \, \rho \, \Tr(\overline{\mathbf{R}}^{\rm MRI}) = \frac{1}{2} \rho \left ( \overline{R}^{\rm MRI}_{rr} + \overline{R}^{\rm MRI}_{\phi\phi}\right ) = \rho \, \overline{v_r^\prime v_r^\prime}\,. \label{eq:eMRI1}
\end{equation}
We can now define the next proportionality constants between the different (MRI) stresses and the MRI energy density:
\begin{align}
    \alpha^{\rm MRI}_{ij} &\equiv \frac{\overline{M}^{\rm MRI}_{ij}}{e_{\rm MRI}}\,, \\
    \beta^{\rm MRI}_{ij} &\equiv \frac{\rho\, \overline{R}^{\rm MRI}_{ij}}{e_{\rm MRI}}\,, \\
    \gamma^{\rm MRI}_{ij} &\equiv \frac{\sqrt{\rho}\, \overline{F}^{\rm MRI}_{ij}}{e_{\rm MRI}}\,, 
\end{align}
whose values can be computed directly from the expressions above
\begin{align}
    \alpha^{\rm MRI}_{rr} = \alpha^{\rm MRI}_{\phi\phi} = -\alpha^{\rm MRI}_{r\phi} &= 4/q-1\,, \label{eq:alphaMRI}\\
    \alpha^{\rm MRI}_{rz} = \alpha^{\rm MRI}_{\phi z} = -\alpha^{\rm MRI}_{z z} &= 0\,,\\
    \beta^{\rm MRI}_{rr} = \beta^{\rm MRI}_{\phi\phi} = \beta^{\rm MRI}_{r\phi} &= 1\,,\\
    \beta^{\rm MRI}_{rz} = \beta^{\rm MRI}_{\phi z} = -\beta^{\rm MRI}_{z z} &= 0\,, \\
    \gamma^{\rm MRI}_{ij} &=0\,. \label{eq:gammaMRI}
\end{align}

For the case with buoyancy ($\hat \omega_{\rm G}\ne0$) the expressions for $k_{\rm MRI}$ and $\gamma_{\rm MRI}$ become more complex, but can still be computed from the mean quantities \citep[see][]{OBERGAULINGER_2009A&A...498..241O}. Similarly, the coefficients $\alpha^{\rm MRI}_{ij}$, $\beta^{\rm MRI}_{ij}$ and $\gamma^{\rm MRI}_{ij}$ will have a different dependence. In this work we focus on the particular case without buoyancy, but a priori nothing prevents to extend this approach to the most general case.

\subsubsection{Parasitic instabilities}
\label{sec:PI}

In the high Reynolds and magnetic Reynolds number regime, the dominant parasitic mode is of the Kelvin-Helmholtz type and develops along the MRI velocity channel \citep{Pessah_2009}. It is characterised by a wavenumber for the fastest growing parasitic mode \citep{Pessah_2010}
\begin{equation}
    k_{\rm PI} = 0.59\,k_{\rm MRI}\,,
\end{equation}
and a corresponding growth rate
\begin{equation}
    \gamma_{\rm PI} = 0.45 \, k_{\rm PI} \, v_0 e^{\gamma_{\rm MRI} t} = \sigma\, k_{\rm MRI} \, v_0 e^{\gamma_{\rm MRI} t}\,,
\end{equation}
with $\sigma=0.27$. Note that, since the growth rate depends on the amplitude of the channel flow, $v_0 e^{\gamma_{\rm MRI} t}$, the PI will grow super-exponentially.
Using the channel mode expressions we can relate $v_0\, e^{\gamma t}$ to $e_{\rm MRI}$ and rewrite the PI growth rate as
\begin{equation}\label{growth_rate_pi}
    \gamma_{\rm PI} = \sigma \, k_{\rm MRI} \, \sqrt{\frac{2\, e_{\rm MRI}}{\rho}}\,.
\end{equation}

Since parasitic modes grow on top of the channel modes, the velocity and magnetic field perturbations can be decomposed as
\begin{equation}
    \mathbf{v}^\prime ={\mathbf{v}^\prime}^{\rm MRI} +{\mathbf{v}^\prime}^{\rm PI}\quad, \qquad
    \mathbf{B}^\prime ={\mathbf{B}^\prime}^{\rm MRI} +{\mathbf{B}^\prime}^{\rm PI}\,.
\end{equation}
The averaged Reynolds stress will then be
\begin{align}
    \overline{R}_{ij} &= \overline{R}_{ij}^{\rm MRI} + \overline{R}_{ij}^{\rm PI} + \overline{v^{\rm MRI}_i v^{\rm PI}_j} +  \overline{v^{\rm PI}_i v^{\rm MRI}_j}\,.  
\end{align}

If we consider that MRI and PI modes are spatially uncorrelated, the last two terms in the previous equation can be dropped. Alternatively, we could absorb these two terms into the definition of $\overline{R}_{ij}^{\rm PI}$, since they are zero when no PI modes are present. In either case we can decompose the averaged Reynolds stress into a component coming from the MRI (already computed in the previous section) and a component coming from the PI. Similar arguments can be made for the Maxwell and Faraday stresses leading to the decomposition
\begin{equation}
 \overline{\mathbf{R}} = \overline{\mathbf{R}}^{\rm MRI} +\overline{\mathbf{R}}^{\rm PI}\quad,\quad     
 \overline{\mathbf{M}} = \overline{\mathbf{M}}^{\rm MRI} +\overline{\mathbf{M}}^{\rm PI}\quad,\quad     
 \overline{\mathbf{F}} = \overline{\mathbf{F}}^{\rm PI}\,.    
\end{equation}

The (kinetic) energy density of the parasitic modes can be computed as
\begin{equation}
    e_{\rm PI} = \frac{1}{2} \, \rho \, \Tr(\overline{\mathbf{R}}^{\rm PI}) = 
    \frac{1}{2} \, \rho \, \left ( \overline{R}^{\rm PI}_{rr} + \overline{R}^{\rm PI}_{\phi\phi} + \overline{R}^{\rm PI}_{zz}  \right )\,.
\end{equation}

The role of the PI is to disrupt the channel flows generating turbulence. To simplify our model we assume that the turbulence generated by the PI is isotropic, meaning that all diagonal components of $\overline{R}^{\rm PI}_{ij}$ are equal. Under this condition
\begin{equation}
    e_{\rm PI} = \frac{3}{2}\rho \, \overline{R}_{zz}\,, \label{eq:ePI}
\end{equation}
where we have used that $\overline{R}^{\rm PI}_{zz}=\overline{R}_{zz}$ because $\overline{R}^{\rm MRI}_{zz}=0$.
Additionally, the isotropy condition allows us to rewrite Eq.~\eqref{eq:eMRI1} in terms of the Reynolds stress, instead of only the MRI components, as
\begin{equation}
    e_{\rm MRI} = \frac{1}{2}\rho \left ( \overline{R}_{rr} + \overline{R}_{\phi\phi} - 2 \overline{R}_{zz}\right )\,. \label{eq:eMRI}
\end{equation}
Eqs.~\eqref{eq:ePI} and \eqref{eq:eMRI} can be used to estimate the energy density in both types of instabilities in numerical simulations, as in the case of Fig.~\ref{fig:stresses}. We have to keep in mind that this estimator is not perfect and its application to numerical simulations may lead to artefacts in some cases (see discussion in  \citet{paper_simulation} for details). 
%

\subsubsection{Closure relation}

Now that we understand that the different stresses depend on $e_{\rm MRI}$ and $e_{\rm PI}$ we can aim at building a phenomenological relation between the two energy densities.
An important assumption we make is that the proportionality coefficients do not depend explicitly on time and position, only through their dependence of mean field variables. This means that the stress tensors will have the same time dependence as the energy densities $e_{\rm MRI}$ and $e_{\rm PI}$ as long as the mean quantities are constant. We hence propose the next closure relation:
\begin{subequations}\label{closure_coeffs}
\begin{align}
	\mean{M}_{ij}(t,\vect{r}) & = \alpha^{\rm MRI}_{ij} \,e_{\rm MRI}(t,\vect{r}) + \alpha^{\rm PI}_{ij} \,e_{\rm PI}(t,\vect{r})\,, \label{Maxwell_coeff}\\
	\mean{R}_{ij}(t,\vect{r}) & = \frac{1}{\mean{\rho}(t, \vect{r})} \left ( \beta^{\rm MRI}_{ij} \,e_{\rm MRI}(t,\vect{r}) + \beta^{\rm PI}_{ij} \,e_{\rm PI}(t,\vect{r}) \right)\,, \\
	\mean{F}_{ij}(t,\vect{r}) & = \frac{\gamma^{\rm PI}_{ij}}{\sqrt{\mean{\rho}(t,\vect{r})}} e_{\rm PI}(t,\vect{r}) \,, 
\end{align}
\end{subequations}
 where the factors involving the mass density are added in order to make the coefficients dimensionless. For the MRI coefficients we use the ones derived for the MRI channel modes, Eqs.~\eqref{eq:alphaMRI}-\eqref{eq:gammaMRI}. We note that the diagonal components of $\beta^{\rm PI}_{ij}$ could be determined from Eq.~\eqref{eq:ePI} and the isotropy condition, resulting in $\beta^{\rm PI}_{ii} = 2/3$. The value of off-diagonal components depends on the correlations between different components of the velocity entering on the averages but are constrained to be $|\beta^{\rm PI}_{ij}| \le 2/3 (i\ne j)$. However, we prefer to keep $\beta^{\rm PI}_{ij}$ as a free parameter of the theory, together with $\alpha^{\rm PI}_{ij}$ and $\gamma^{\rm PI}_{ij}$. These coefficients can be calibrated using numerical simulations and their values are discussed in the next sections. 

\subsubsection{Evolution equations of the MRI and PI energy densities}

The last step is to obtain evolution equations for the the PI and MRI energy densities. From their definitions, Eqs.~\eqref{eq:ePI} and \eqref{eq:eMRI}, the mean-field continuity equation and Eq.~\eqref{eq:ogilvieR} (neglecting the stretching terms) we obtain equations of the form
\begin{align}
    \partial_t e_{\rm MRI }+ \partial_i (\overline{v}_i\,e_{\rm MRI}) &= S^{\rm (MRI)}\,,\\
    \partial_t e_{\rm PI }+ \partial_i (\overline{v}_i\,e_{\rm PI}) &= S^{\rm (PI)}\,,
\end{align} 
i.e. a set of balance laws for both energy densities, with some complicated sources that include both the effect of the MRI and the PI. The stretching terms could be kept and expressed in terms of $e^{\rm MRI}$ and $e^{\rm PI}$ using the coefficients $\beta^{\rm MRI}_{ij}$ and $\beta^{\rm PI}_{ij}$. However, this adds unnecessary complication for the purpose of this paper (stretching is irrelevant for the simulations that we discuss in the next sections) and would make the equations non-conservative.

A closed and simple form for the source terms can be justified taking into account some information we derived in the last few subsections, namely: i) the amplitude of MRI channel flows grows exponentially with a growth rate $\gamma_{\rm MRI}$ and hence $e_{\rm MRI}$ grows with a rate of $2\gamma_{\rm MRI}$ during the initial phase; ii) The same happens for $e_{\rm PI}$, but with a growth rate $2\gamma_{\rm PI}$; iii) The PI draws energy from $e_{\rm MRI}$ whenever both energy densities are comparable, quenching the growth of the MRI channels and leading to saturation; iv) Finally, the PI generates turbulence in which larger vortices are broken in smaller ones in a turbulent cascade until the physical dissipation scale is reached where dissipation occurs. At this small scale,  kinetic (and magnetic) energy is transformed into thermal energy \citep[see e.g.][chapter III]{LANDAU198795}. Taking all these considerations into account we propose the next set of equations for the evolution of MRI and PI energy densities:
\begin{eqnarray}
\label{energy_dens_ev}
    \partial_t e_{\rm MRI} +\partial_i(\mean{v}_i e_{\rm MRI}) &=& 2\,\gamma_{\rm MRI}\,e_{\rm MRI}-2\,\gamma_{\rm PI}\,e_{\rm PI}  \,,
    \\ \label{energy_dens_ev_PI}
    \partial_t e_{\rm PI} +\partial_i(\mean{v}_i e_{\rm PI}) &=& 2\,\gamma_{\rm PI}\,e_{\rm PI}-S_{\rm TD} \,.
\end{eqnarray}

The growth rates $\gamma_{\rm MRI}$ and $\gamma_{\rm PI}$ are given in Eqs.~\eqref{growth_rate_mri} and \eqref{growth_rate_pi}, respectively. The set of equations represents the energy flow in systems unstable to MRI. Firstly, the term $2 \, \gamma_{\rm MRI} e_{\rm MRI}$ draws energy from the averaged quantities to generate MRI channel flows, increasing the MRI energy. Secondly, the term $2\,\gamma_{\rm PI}\,e_{\rm PI}$ acts as an energy sink for $e_{\rm MRI}$; this energy is transferred in a conservative way to the equation for $e_{\rm PI}$, where it acts as a source. Finally, the quantity $S_{\rm TD}$, representing the turbulent energy dissipation at the end of the Kolgomorov cascade, transfers energy back to the averaged quantities in terms of thermal (internal) energy.

In order to solve Eq.~\eqref{energy_dens_ev_PI} one also needs an expression for the turbulent energy density dissipation term, $S_{\rm TD}$. In Section 33 of \cite{LANDAU198795} there is an estimate for the energy dissipation rate per unit mass, $\epsilon = S_{\rm TD}/\rho$. It is shown that
\begin{equation}\label{eq:v3lambda}
    \epsilon \propto v_{\lambda}^3/\lambda\,,
\end{equation}
where $\lambda$ corresponds to the size of the turbulent eddy (the order of magnitude of the distances over which changes in velocity can be appreciated) and $v_{\lambda}$ is the speed of the turbulence in spatial scales $\sim \lambda$. In the inertial range of scales (Kolmogorov cascade) the flux of energy to smaller scales should be constant across the range of wave vectors and then $\epsilon$ should not depend on $\lambda$ \citep{LANDAU198795}. Since the turbulence we are now considering is developed by the PI, $v_{\lambda}$ could be estimated from $e_{\rm PI}$ as
\begin{equation}\label{vel_pi}
    v_{\lambda} \propto \sqrt{\frac{e_{\rm PI}}{\rho}}\,.
\end{equation}

It is possible to demonstrate that this velocity corresponds to the scale with the largest eddy size fitting within the filter size used to perform the averages. According to Kolgomorov's theory, the kinetic energy spectrum scales as $E(k) \propto \epsilon^{2/3} k^{-5/3}$. If we compute the energy in the inertial range of $k$ fitting within a certain volume (e.g., the volume used to average the MHD equations) we obtain: 

\begin{equation}
    E_{\rm V} = \int_{k_{\rm min}}^{k_{\rm max}} E(k)dk \propto \int_{k_{\rm min}}^{k_{\rm max}} \epsilon^{2/3}k^{-5/3} dk \,,
\end{equation}
where the integral is computed between $k_{\rm min}=1/\lambda$ (corresponding to the size of the volume considered, which corresponds to the size $\lambda$ of the largest eddy fitting the volume) and $k_{\rm max}$ (size of the smallest scale, i.e. the dissipation scale). Since we are considering the case with large Reynolds number, $k_{\rm max} \gg k_{\rm min}$, and the integral results in
\begin{equation}
    E_{\rm V}  \propto \epsilon^{2/3}k_{\rm min}^{-2/3} = (\epsilon\lambda)^{2/3} \propto v_\lambda^2\,,
\end{equation}
where we have used Eq.~\eqref{eq:v3lambda} to express the integral in terms of the velocity.

\begin{table*}
    \centering
\begin{tabular}{|cccccccc|}
\hline
    NAME & $\mean{B}_{0z}$ [$10^{13}$ G] & Resolution $(r\times \phi \times z)$ & Box size [km] & $\lambda_{\rm MRI}$ [km] & Zones per channel \\
    \hline
    \hline
    MRI-L1 & $4.6$ & $60\times 240 \times 60$ & $1\times 4 \times 1$ & $0.333$ & 20   \\
    MRI-M1 & $4.6$ & $76\times 304 \times 76$ & $1\times 4 \times 1$ & $0.333$ & 25  \\
    MRI-H1 & $4.6$ & $100\times 400 \times 100$ & $1\times 4 \times 1$ & $0.333$ & 33   \\
    MRI-H2 & $3.45$ & $100\times 400 \times 100$ & $1\times 4 \times 1$  & $0.25$ & 25   \\
    MRI-H3 & $2.76$ & $100\times 400 \times 100$ & $1\times 4 \times 1$ & $0.2$ & 20  \\
\hline
\end{tabular}
    \caption{This table reports the different MRI simulations from \citet{Rembiasz-2016} used to test the sub-grid models discussed in this work. The simulations differ in the numerical resolution and in  the initial magnetic-field strength. Note that $\lambda_{\rm MRI}$ is not uniform throughout the computational domain, but varies by $\approx 20 \%$.}
    \label{tab:table_sim}
\end{table*}

This means that the bigger scales contribute more to the value of $e_{\rm PI}$, where $e_{\rm PI} = E_{\rm V}/\rm{V}$, and justifies the use of Eq.~\eqref{vel_pi}. In this case, we interpret that $e_{\rm PI}$ is essentially related to the minimum value between the minimum wavelength of the modes of the MRI and the size of the filter. If all of the excited modes of the instability are resolved (i.e., the minimum wavelength of the instability is larger than the filter size), $\lambda$ will be equal to the size of the filter, $\Delta_f$. However, if the scales of the instability are not resolved, the PI should be described by the scale of the instability, which is $\lambda_{\rm MRI}$. Therefore we assume
\begin{equation}
    \lambda = \min[\Delta,\lambda_{\rm MRI}]\,.
\end{equation}

Eqs.~\eqref{eq:v3lambda} and \eqref{vel_pi} allow us to give an expression for the energy dissipation rate per unit mass
\begin{equation}\label{energy_dis_per_mass}
    \epsilon \propto \frac{1}{\lambda}\Big(\frac{e_{\rm PI}}{\rho}\Big)^{3/2}\,,
\end{equation}
which can be used to estimate the turbulent energy density dissipation rate
\begin{equation}\label{std}
    S_{\rm TD} = \rho \epsilon = C\frac{e_{\rm PI}^{3/2}}{\sqrt{\rho}\lambda}\,,
\end{equation}
where $C$ is a dimensionless constant.

The interpretation of this constant $C$ can be understood by studying Eq.~\eqref{energy_dens_ev_PI} at the saturation point $t_{\rm sat}$, i.e. when $\partial_t e_{\rm PI} = 0$. At this maximum, the right hand side of Eq.~\eqref{energy_dens_ev_PI} vanishes and it is possible to compute the ratio of PI to MRI energy 
\begin{equation}
    \frac{e_{\rm PI}(t_{\rm sat})}{e_{\rm MRI}(t_{\rm sat})} = \frac{8 \sigma^2}{C^2} \lambda^2 k^2_{\rm MRI} = \frac{8 \sigma^2}{C^2} \frac{\Omega^2\lambda^2 \rho}{\overline{B}_z^2} \left (q-\frac{q^2}{4} \right).
\end{equation}
Therefore, the value of $C$ is directly related to the ratio of PI to MRI energy in the saturated state.

\begin{figure*}
    \centering
    \includegraphics[width=\textwidth]{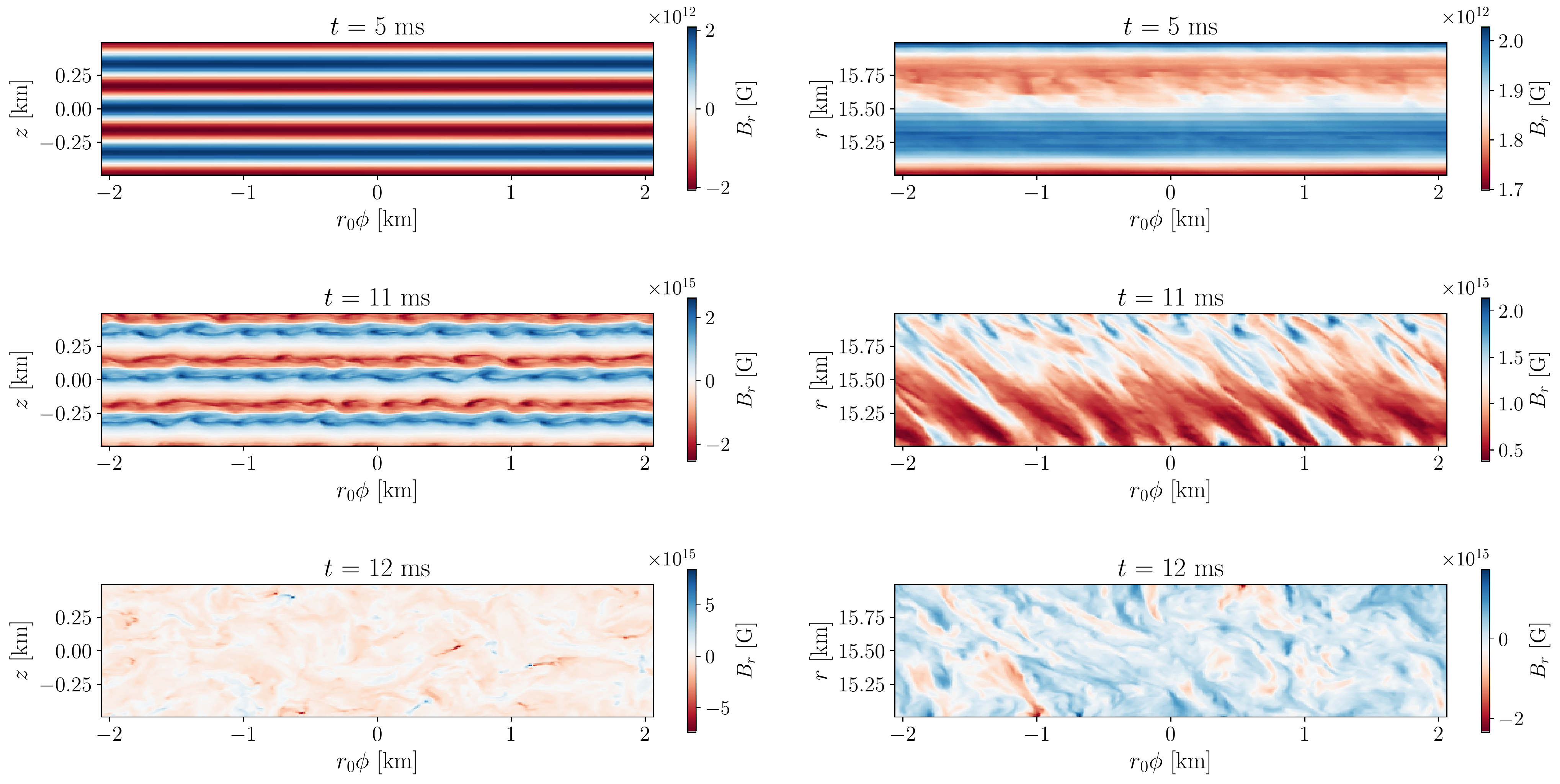}
    \caption{Cuts of the radial component of the magnetic field for the simulation MRI-H1, at the $z\phi$ plane (left column) and at the $r\phi$ plane (right column). The rows show different times of the simulation, $t={5,11,12}$ ms. The MRI sets in from a layered  distribution of $B_r$ ($t=5$ ms),  grows exponentially ($t=11$ ms) until it saturates and reaches an isotropic and turbulent configuration ($t=12$ ms).}
    \label{fig:br_cuts}
\end{figure*}

\section{Results}
\label{sect: sec4}

\subsection{Box simulations}
\label{simulations}

We test the different sub-grid models discussed in the previous sections using a subset of the three-dimensional, semi-global MRI simulations of \cite{Rembiasz-2016}. The simulation domain is a section of a cylindrical annulus with a size of $1 \, \mathrm{km} \times 4 \, \mathrm{km} \times 1 \, \mathrm{km}$ in the radial direction, $\phi$ (i.e., rotational direction), and $z$ direction, respectively. The five models we use are summarized in Table~\ref{tab:table_sim} and they correspond to simulations with different grid resolutions and initial magnetic-field strength. The dynamics of the plasma is governed by the Newtonian visco-resistive MHD equations and the simulations were performed using the \texttt{AENUS} code \citep{Obergaulinger-2008}. While the resistivity and the shear and bulk  viscosities of the simulations  are non-zero, they are sufficiently small so as not  affect the growth rates of the MRI and the KHI, the latter acting as the PI terminating the growth of the MRI. Therefore, all the three sub-grid models described in Section~\ref{sect: sec3} can in principle be used with these simulations.

The total pressure consists of a polytropic part and a thermal part \citep{Dimmelmeier2001GeneralRC}, $P = P_{\rm p}+P_{\rm th}$, where $P_{\rm p}$ reproduces the pressure exerted by a degenerate electron gas while $P_{\rm th}$ models a finite-temperature correction. The total pressure is
\begin{equation}
    P = P_{\rm p}+P_{\rm th} = K\rho^{\gamma}+(\gamma_{\rm th}-1)\rho\Big(\varepsilon-\rho^{\gamma-1}\frac{K}{\gamma-1}\Big)\,,
\end{equation}
where $\varepsilon$ is the specific internal energy, $K = 4.8974894 \times 10^{14}$ (in cgs units), $\gamma = 1.31$ and $\gamma_{\rm th} = 1.5$. This is a good representation of the EOS at the sub-nuclear densities considered in the simulations.

\begin{figure*}
    \centering
    \includegraphics[width=\textwidth]{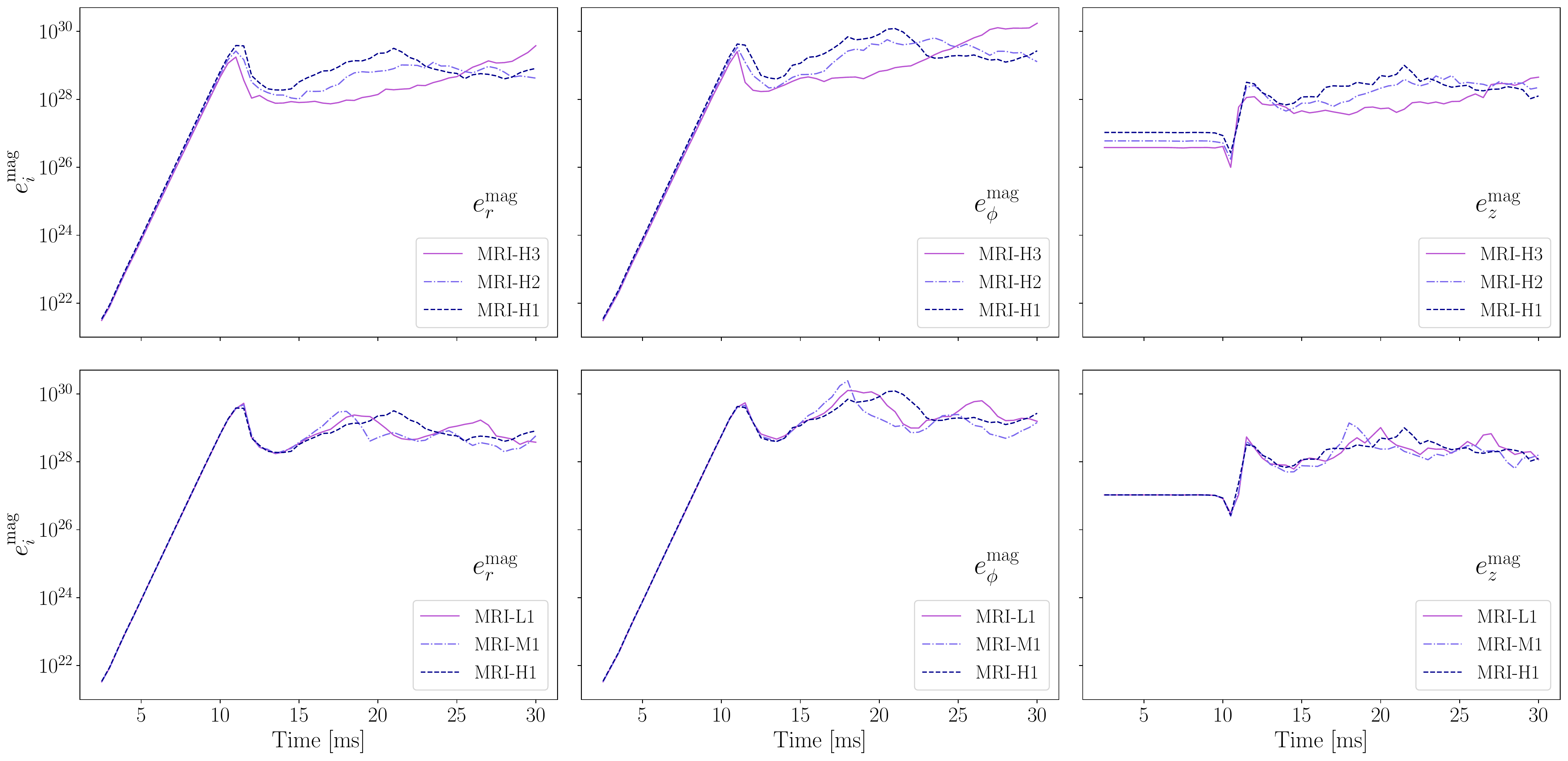}
    \caption{Time evolution of the averaged magnetic energy density components, radial (left), azimuthal (centre), and vertical (right), in cgs units. The top panels correspond to simulations with the same resolution but different initial magnetic field strength. The bottom panels depict results for different resolutions but the same value of the initial magnetic field (cf.~Table \ref{tab:table_sim}).}
    \label{fig:mag_energy}
\end{figure*}

The initial conditions for the simulations approximate the equatorial layer of an MRI-unstable surface layer of a proto-neutron star, which are similar to the post-merger configuration of a BNS merger. We use a  differential rotation profile $\Omega (r) \propto r^{-q}$ (see Eq.~\eqref{ang_velocity}), with $\Omega_0 = 1824$~$\rm{s}^{-1}$, a characteristic radius (center of the box) $r_0=15.5$~km and a rotational shear $q=1.25$. The structure of the layer was chosen such as to maintain hydrostatic equilibrium, i.e., balance between the gravitational force of the star, the gas pressure, and the centrifugal force, and marginal convective stability, i.e., a flat pseudo-entropy ($s = P / P_{\rm p}$) profile. This results in a central density of $\rho_0=1.83\times10^{-15}$, that corresponds to $2.47\times 10^{13}$~g~cm$^{-3}$, typical of the regions developing MRI in proto-neutron stars and BNS mergers \citep{Rembiasz-2016}. The initial magnetic field has only a uniform vertical component $B_{0z}$. Compared to the internal or rotational energy of the gas, the magnetic field is weak. In all 5 models, the most unstable MRI channel modes have a wave length of $\lambda_{\rm MRI} \leq 0.333\, \mathrm{km}$. This scale is resolved by at least 20 grid cells, which means that the MRI growth rate is numerically close to convergence \citep[see][]{Rembiasz2016}.

\subsubsection{Global quantities.}

The onset, growth and termination of the MRI can be observed by monitoring the evolution of different quantities. Fig.~\ref{fig:br_cuts} displays cuts on the $z\phi$ plane (left) and on the $r\phi$ plane (right) of the radial component of the magnetic field, $B_r$, for the simulation MRI-H1 at three different times. Shortly after the start of the simulation ($5$~ms, upper panels) channel modes appear triggered by the applied initial perturbations and the magnetic field exponentially grows from $10^{12}$ G to $10^{15}$ G (middle panels), point at which parasitic KH instabilities start to be visible. Eventually the field saturates and a turbulent configuration is reached after $t \approx 12$ ms (bottom panels). The upper right panel of Fig.~\ref{fig:br_cuts} shows that the magnetic field (and similarly all other variables) is not completely smooth in the radial direction. As mentioned before, this is an artefact of the boundary conditions discussed in \cite{paper_simulation}. The artefact is related to the use of the so-called shearing disc boundary condition in the radial direction; this approximate treatment of the radial boundaries is necessary because there is radial dependence of background quantities that does not allow for the use of periodic-like (shearing box) conditions. This limits our ability to compute $e_{\rm PI}$ from the simulation which, in turn, affects the computation of the coefficients for the MInIT sub-grid model
(see discussion in Sections~\ref{sec:PI}, \ref{sec:PICalib} and \ref{sec:OPTC}).

\begin{figure*}
    \centering
    \includegraphics[width=\textwidth]{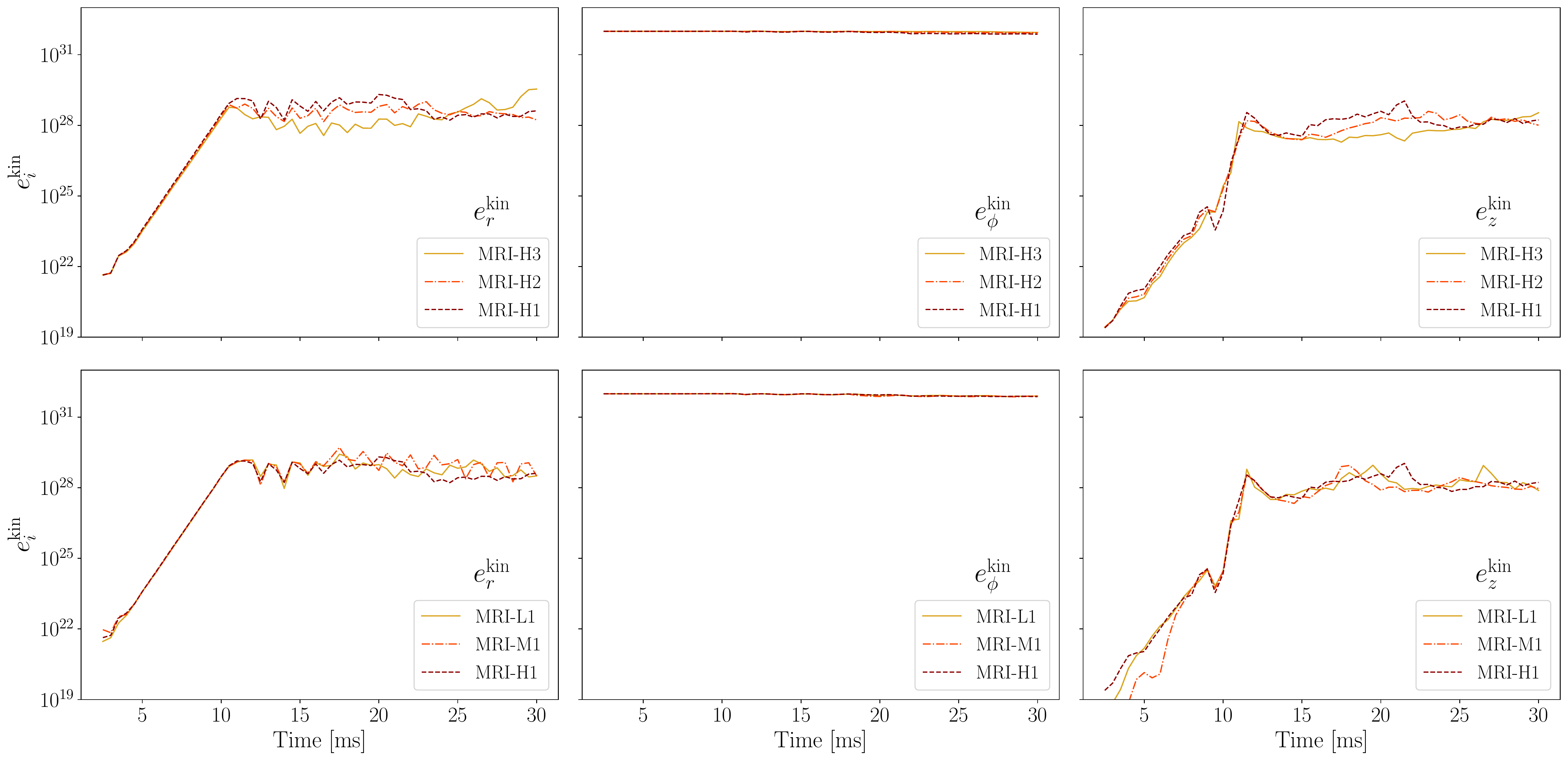}
    \caption{Time evolution of the averaged kinetic energy density components, in cgs units. As in Fig.~\ref{fig:mag_energy}, we display from left to right the radial, azimuthal and vertical components. Again, the top panels correspond to simulations with different initial magnetic field strength and the bottom panels to simulations with different grid resolution.}
    \label{fig:kin_energy}
\end{figure*}

Fig.~\ref{fig:mag_energy} shows the time evolution of the contribution of each component to the total averaged magnetic energy density 
\begin{equation}
    \overline{e}^{\rm mag}_i = \frac{\overline{B_i^2}}{2}\,,
\end{equation}
for different resolutions and initial magnetic fields. No important differences are found between the simulations. All cases plotted show that the MRI grows at the same rate and saturates at roughly the same level. Moreover, the amplification of the magnetic field is similar for the three components of the magnetic field. Note that the growth of the vertical component, shown in the right column of the figure, is shorter because the initial magnetic field points in this direction. 

\begin{figure*}
    \centering
    \includegraphics[width=\textwidth]{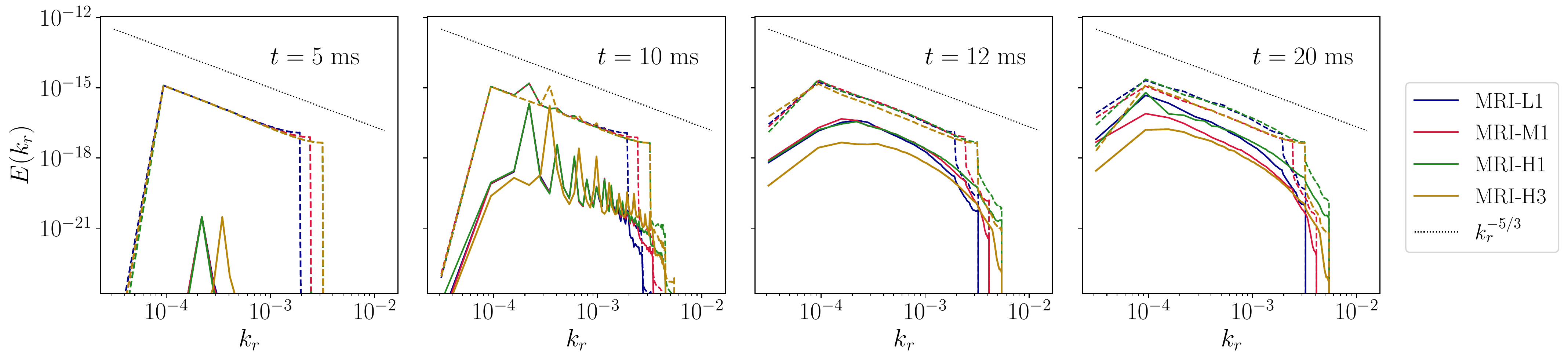}
    \caption{Spectra of the magnetic (solid lines) and kinetic (dashed lines) energies for different MRI simulations. As time increases, the magnetic energy tends to even the kinetic energy, specially at smaller scales, where equipartition is reached after $t\approx 12$ ms.}
    \label{fig:spectra}
\end{figure*}

Furthermore, Fig.~\ref{fig:kin_energy} shows the time evolution of the contribution of each component to the total averaged kinetic energy for the same simulations shown in Fig.~\ref{fig:mag_energy}. The expression for each component is 
\begin{equation}
    \overline{e}^{\rm kin}_i = \frac{1}{2}\overline{\rho v_i^2} \,.
\end{equation}
The only non-vanishing component of the initial velocity field is the azimuthal one. This component remains nearly constant during the whole simulation (middle panel) whereas the radial and vertical components grow exponentially until they saturate again at $t \approx 12$ ms. No remarkable differences are found among the 5 simulations. 

A way to see how the magnetic field is amplified during the MRI is via the energy spectra. This is shown in Fig.~\ref{fig:spectra} at three representative times. Equipartition between kinetic and magnetic energies is reached faster at smaller scales (higher $k$). Towards the end of the simulations, at $t \approx 20$ ms, the two energies tend towards equipartition at all scales. The simulation with higher resolution, MRI-H1, reaches equipartition at small scales earlier than the rest, while the simulation with the smallest initial magnetic field, MRI-H3, is the slowest to reach a high value of the magnetic field at large scales. 

\subsection{Determination of the $\alpha$-dynamo coefficient}

\begin{figure}
    \includegraphics[width=0.47\textwidth]{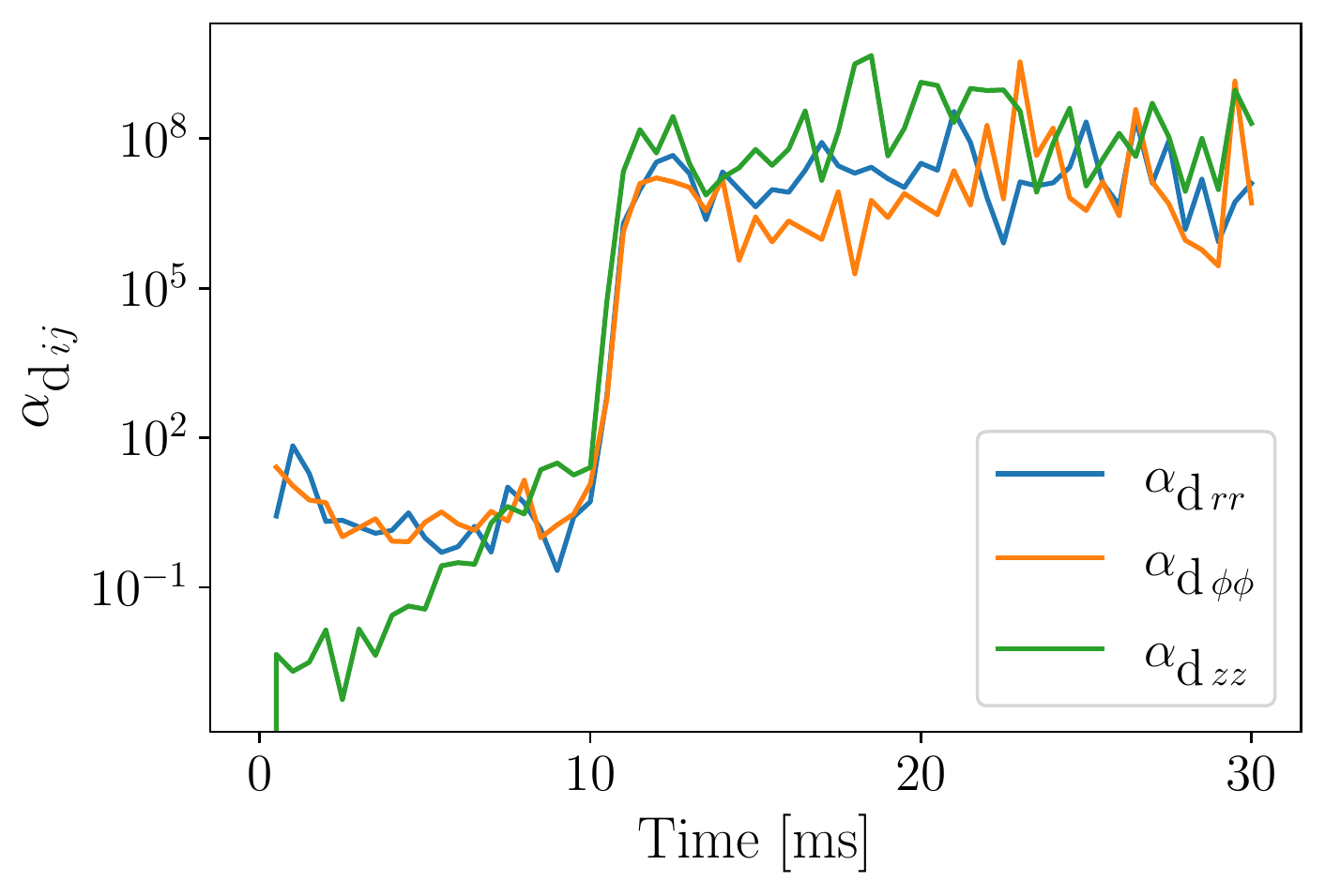} 
    \caption{Time evolution of the diagonal components of the $\alpha$-dynamo coefficient, computed from simulation MRI-H1. After an exponential growth, all components reach a similar value at saturation.}
    \label{fig:alpha_beta_dyn}
\end{figure}

 Our numerical simulations can also be employed  to obtain the dynamo coefficients $\alpha_{\rm d}$ and $\beta_{\rm d}$. Here we use Eq.~\eqref{final_close_alpha} to compute $\alpha_{\rm d}$ assuming that this coefficient is constant inside the simulation box (remember the discussion in Section~\ref{sect:dynamos} to neglect the computation of $\beta_{\rm d}$). In Fig.~\ref{fig:alpha_beta_dyn} we show the time evolution of the $\alpha_{\rm d}$ coefficient, computed as the average over the azimuthal direction and evaluated at the center of the $rz$ plane. There is an initial stage at which all diagonal components are nearly constant, except for $\alpha_{{\rm d}\,zz}$, which starts at a much lower initial value. At $t\approx 10$ ms, all components grow exponentially for a short time. This is followed by a saturation phase where all components reach a similar and almost constant level. This result supports the use of the $\alpha_{\rm d}$ coefficient as a scalar. However,  setting a constant value for this coefficient {\it at all times} of a simulation could lead to wrong results, specially during the phase in which the exponential growth occurs. This is the reason why we search for a sub-grid model able to deal with temporal evolutions and capture all different stages of the development of turbulence.
 
\subsection{{Averaging procedure}}
\label{sec:filter}

One important aspect to test the different sub-grid models is to have a proper definition of what we mean by average, a procedure that is performed using a filter. There are different filters that can be used to obtain the mean component of a turbulent field. For example, \cite{paper_gradient_test} present the Gaussian filter, which is used in the formal development of the gradient model. It has nice mathematical properties, but it is not used in the a-priori tests. Alternatively, in the a-priori comparison the authors perform an average over $S_f^3$ cells, using the so-called box filter. In this paper we also use a box filter. Each filter size is labeled by the quantity $S_f = \Delta_f / \Delta$, where $\Delta$ is the size of the cell of the direct numerical simulation and $\Delta_f$ is the filter size. In the case of the box filter, $\Delta_f$ corresponds to the size of the box, which contains  $S_f$ cells per dimension. Thus, a filtered quantity is the mean value in a domain with size $\Delta_f$ for each direction. Note that even though our simulations used cylindrical coordinates, the filtering operation will be performed over a domain with equal length per dimension, since we need a unique value for the filter size, $\Delta_f$. In fact, in our case the cell size $\Delta$ is different for each direction, namely $\Delta r = \Delta z \neq r \Delta \phi$, and therefore the filter size will also be different for each direction, $\Delta_f^i$, giving the same value of $S_f$ in the three directions. 

The box filter is characterized by the following normalized kernel \citep{paper_gradient_test}: 
\begin{equation}\label{box_filter}
    F_i(|r_i-r^{\prime}_i|) = \left \{ \begin{array}{cc}
        1/\Delta_f & \mbox{ if } \,\, |r_i-r^{\prime}_i| \leq  \Delta_f^i/2 \,,\\
         0 & \mbox{ if } \,\, |r_i-r^{\prime}_i| > \Delta_f^i/2 \,,
    \end{array} \right.
\end{equation}
for each dimension. The three-dimensional kernel will be
\begin{equation}\label{box_3d}
    F (|\textbf{r}-\textbf{r}^{\prime}|) = \prod_i^3 F_i(|r_i-r^{\prime}_i|)\,.
\end{equation}

As mentioned before, our simulations use cylindrical coordinates $(r, \phi, z)$ and therefore $\Delta$ is different for each side of a numerical cell. However, this is not an issue for the box filter where the filter has the same shape than the grid cell $\Delta$. In fact, box-filtered data can be regarded as data from a simulation with lower resolution, i.e.~using effectively bigger cells than the actual ones in the simulation.

\subsection{Numerical implementation and calibration of the MInIT model}

\subsubsection{Energy density evolution equations in the MInIT model}

In order to apply the MInIT model one has to integrate numerically Eqs.~\eqref{energy_dens_ev} and \eqref{energy_dens_ev_PI} in time, starting with appropriate initial values at $t=0$, $e_{\rm MRI}(0)$ and $e_{\rm PI}(0)$ (initial conditions are discussed in Section~\ref{sec:OPTC}). For the box simulation considered in this work, the average velocity $\overline{\mathbf{v}}$ only has non-zero $\phi$ component. In principle, the advective term in the $\phi$ direction should therefore be considered. However, since the simulation has periodic boundary conditions in the $\phi$ direction, any spatial average over the whole simulation box (or involving averages over $\phi$) will be independent of whether this advection was actually performed or not. Even if we use a filter of size $\Delta_f$, the result will be independent of this advection term, at least in a statistical sense, as long as we construct the final quantities as averages of filtered quantities at different places in the whole box. Therefore, we will not consider the advection term for the calibration and tests performed in this work.

The second consideration is the calculation of the coefficients in the right-hand-side of  Eqs.~\eqref{energy_dens_ev} and \eqref{energy_dens_ev_PI}. Those coefficients depend on the mean quantities $\rho$, $\overline{B}$ and $\overline{v}$, the latter through the values of $\Omega$ and $q$. There are small differences in the values of these quantities across the box, and even within the filter region.

Finally, we need a numerical procedure to integrate numerically  Eqs.~\eqref{energy_dens_ev} and \eqref{energy_dens_ev_PI}. This system of partial differential equations is in general stiff because it involves the exponential and super-exponential growth of the quantities. Therefore, care has to be taken in the time integration. We use the \textit{Strang splitting} method \citep{Strang:1968} to solve them.

Given the initial conditions of the box simulation, which provide the average values of $\rho$, $\overline{B}$ and $\overline{v}$, and all  calibrated coefficients of the MInIT model (see next two sections), the integration of the equations directly provide a model for the whole simulation, as long as the average values do not change, which is approximately true in our box simulations. The result the model yields can then be compared with that from the numerical simulation to assess its accuracy.

\subsubsection{Calibration of the PI coefficients of the MInIT model}
\label{sec:PICalib}

The free coefficients $\alpha^{\rm PI}_{ij}$, $\beta^{\rm PI}_{ij}$ and $\gamma^{\rm PI}_{ij}$ appearing in the closure relations of the MInIT model can be computed from Eqs.~\eqref{closure_coeffs} by averaging the stresses in space and time over the whole box of the simulation and a representative simulation time. In order to avoid initial transients that appear in some tensor components during the growth phase (an artefact of the boundary conditions discussed in \cite{paper_simulation}) we only average at times after saturation ($t \gtrsim 12$ ms). The main problem at early times is that $e_{\rm PI}$ has a small value and any small boundary effect produces a large uncertainty in its estimation, introducing a large error in the coefficients. After saturation  $e_{\rm PI}$ becomes comparable to  $e_{\rm MRI}$ and the small boundary errors become negligible. 

\begin{table}
    \centering
\begin{tabular}{|c c c c|}
\hline
     & $\alpha^{\rm PI}_{ij}$ & $\beta^{\rm PI}_{ij}$ & $\gamma^{\rm PI}_{ij}$   \\
    \hline
    \hline
   $ rr $ & $0.5 \pm 1.2$ & $0.08\pm 0.55$ & -   \\
    \hline
    $\phi \phi$ & $7 \pm 3 $ & $1.2 \pm 0.8 $ & -   \\
    \hline
    $zz$ & $0.8 \pm 0.4$ & $0.7 \pm 0.3$ & -   \\
    \hline
    $r\phi$ & $-1.4 \pm 1.5$ & $-0.8 \pm 0.6$ & $0.10 \pm 0.81$ \\
    \hline
    $rz$ & $0.06 \pm 0.34$ & $0.03 \pm 0.18 $ & $0.02 \pm 0.29$ \\
    \hline
    $\phi z$ & $-0.1 \pm 0.4$ & $0.07 \pm 0.26$ & $-0.10 \pm 0.51$ \\
    \hline 
\end{tabular}
    \caption{Numerical estimation of the PI coefficients of the closure relations of the MInIT model. The uncertainties (standard deviation) arise from both the time and the spatial averages of the stress tensors used to calculate the coefficients. Simulation MRI-H1 was used to compute the coefficients reported here. Statistically similar coefficients were obtained when used other simulations.}
    \label{tab:table_coeffs}
\end{table}
 
The estimated values of the coefficients that we obtain are reported in Table~\ref{tab:table_coeffs}. We find that the diagonal components of $\alpha^{\rm PI}_{ij}$ and $\beta^{\rm PI}_{ij}$ are all positive, with $\alpha^{\rm PI}_{rr}$ and $\beta_{\rm rr}$ being compatible with zero. The $r\phi$ component for both the Maxwell and Reynolds stresses is consistent with a non-zero value, as expected, since it is responsible for the transport of angular momentum in the radial direction. However, the other non-diagonal values are much smaller and consistent with zero. Regarding the assumption of isotropy in the PI, the expectation is to have $\beta^{\rm PI}_{ii} = 2/3$ and $|\beta^{\rm PI}_{ij}| \le 2/3$ for $i\ne j$ (see Section~\ref{sec:PI}). All coefficients estimated are compatible with these predictions within 2-$\sigma$ uncertainties. We observe significant differences among the values of different off-diagonal terms: while the $rz$ and $\phi z$ components are practically zero, the $r\phi$ is close to the upper limit of $2/3$ and negative. This indicated strong anti-correlations between the $r$ and $\phi$ components of $\mathbf{v}'$ and at the same time low correlation with the $z$ component. This is understandable if the PI (of KH type) develops on top of the channel flows, in a vertical plane forming an angle of $\pi/4$ with the radial direction.
This breaks the assumption of isotropy in some sense by establishing a preferred directions. However, it does not affect the assumptions in Section~\ref{sec:PI}.

\begin{figure*}
\centering
   \includegraphics[width=0.8\textwidth]{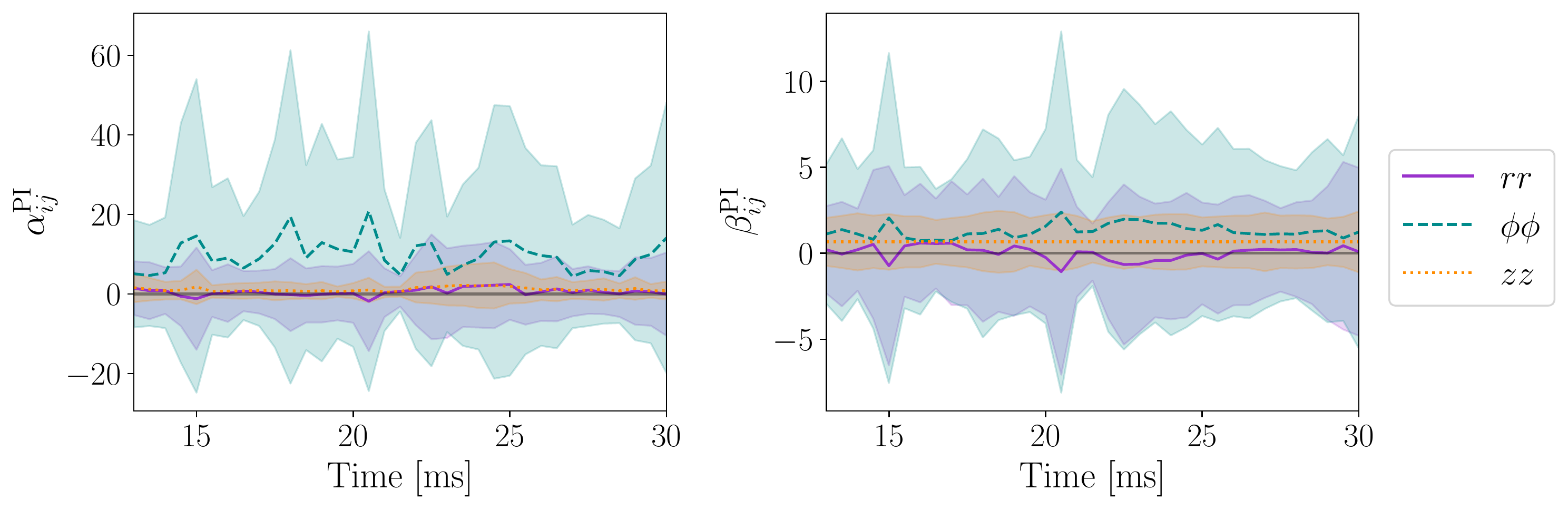}
    \caption{Time evolution of the diagonal components of the $\alpha^{\rm PI}$ and $\beta^{\rm PI}$   coefficients. The shadows represent the standard deviation that arises from the average over the whole simulation box. Note that the values of each component are consistently time-independent.}
   \label{fig:coeffs-diag}
\end{figure*}
 
\begin{figure*}
\centering
   \includegraphics[width=\textwidth]{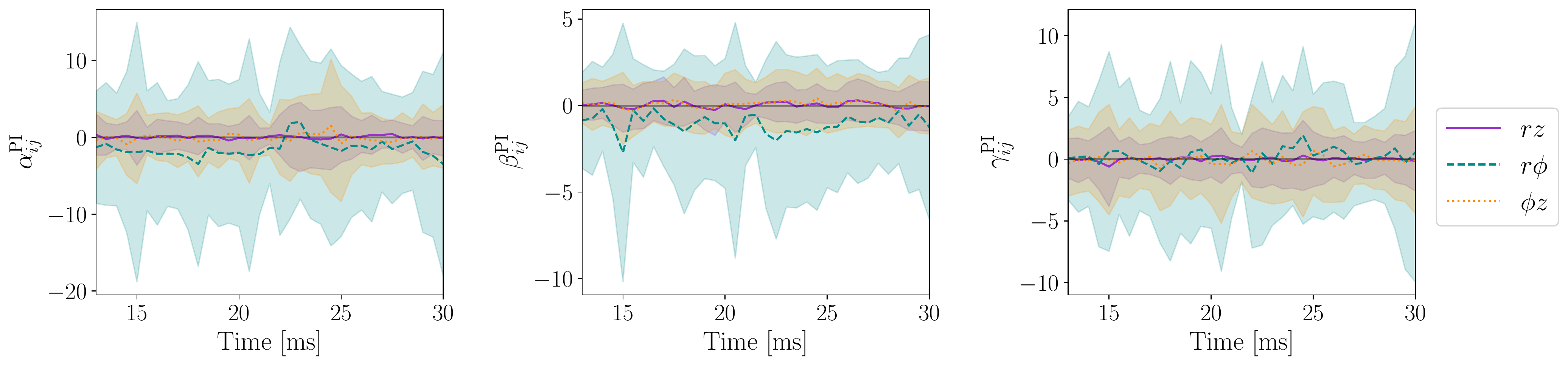}
    \caption{Time evolution of the non-diagonal components of the  $\alpha^{\rm PI}$, $\beta^{\rm PI}$ and $\gamma^{\rm PI}$ coefficients. As in Fig.~\ref{fig:coeffs-diag}, the shadows represent the standard deviation that arises from the average over the whole simulation.}
   \label{fig:coeffs-non-diag}
\end{figure*} 
 
By averaging over the whole box and time (for $t\gtrsim 12$ ms) of the simulation, we are assuming that there are no statistical spatial or temporal variations in the averaging domain and all samples (points and times) are representative of the same quantity we want to measure. In the spatial case, the radial variations of the initial conditions, which are preserved in the averaged quantities during the simulation, are sufficiently small to be neglected. In order to understand the temporal behaviour we show in Figs.~\ref{fig:coeffs-diag} and Fig.~\ref{fig:coeffs-non-diag} the diagonal and non-diagonal components of the coefficients, respectively, computed only from spatial averages, as a function of time. The shaded regions in these figures represent the standard deviation that arises from the average in space of the stresses over the whole box. Comparing with the values from Table~\ref{tab:table_coeffs}, one can see that the change of the  coefficients is larger in space than in time, since the uncertainty from Table~\ref{tab:table_coeffs} is smaller. Indeed, the coefficients oscillate in time around an approximately constant value. 

\subsubsection{Optimization of the C parameter of the MInIT model}
\label{sec:OPTC}

After fixing the PI coefficients of the closure model, we still have three free parameters that need to be fixed, the dimensionless constant $C$, Eq.~\eqref{std}. and the initial values of the energy densities,  $e_{\rm MRI}(0)$ and $e_{\rm PI}(0)$. Of these three parameters, only $C$ is truly a free parameter of the MInIT model. The other two depend on the particular physical system. One could in principle take those two values directly from the simulation. However, the boundary-condition errors mentioned in the previous section introduce large uncertainties in $e_{\rm MRI}(0)$ and $e_{\rm PI}(0)$, which are in general small quantities. Hence, we keep these two quantities as free parameters to be fitted from the simulation. Instead of $e_{\rm PI}(0)$, we use the ratio $K_0 = e_{\rm PI}(0)/e_{\rm MRI}(0)$ as free parameter. The initial value of $e_{\rm MRI}$ determines the time at which saturation is reached, $t_{\rm sat}$. The ratio $K_0$ gives the maximum value of $e_{\rm MRI}$ at $t = t_{\rm sat}$, and the value of $C$ determines the energy density attained at the saturation regime. 

\begin{figure}
    \includegraphics[width =0.9\linewidth]{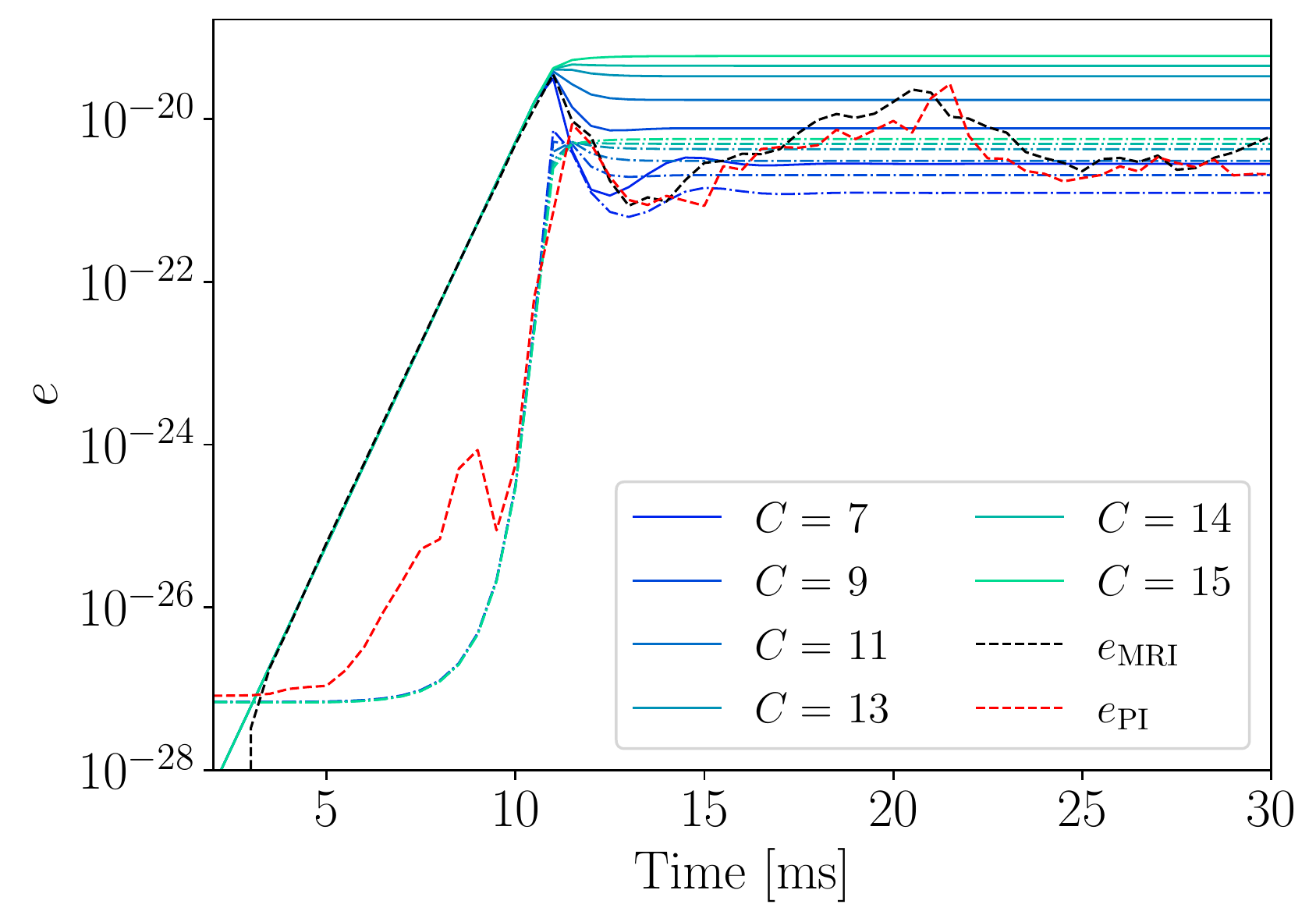}
    \caption{Time evolution of $e_{\rm MRI}$ (solid lines) and $e_{\rm PI}$ (dash-dotted lines) using the MInIT model for different values of the free parameter $C$. The black and red dashed lines show the data from the simulation MRI-H1. We show results only for $K_0 = 1000$ and $e_{\rm MRI}(0) = 7\times 10^{-31}$ as no significant differences are found for other choices.}
    \label{fig:energy_ev}
\end{figure}

In practice, out of the three parameters only $C$ produces changes in the model outcome. We have found that there are no significant changes in the time evolution of $e_{\rm MRI}$ and $e_{\rm PI}$ for values of $K_0$ and $e_{\rm MRI}(0)$ sampled in the range $K_0\in[10, 1000]$ and $e_{\rm MRI}(0)\in[2\times 10^{-31},5\times 10^{-30}]$. Thus, in practice it is sufficient to use sufficiently small values for these initial quantities (e.g.~$K_0 = 1000$ and $e_{\rm MRI}(0) = 7\times 10^{-31}$) since the energy densities grow exponentially and super-exponentially during the development of the instability. Fig.~\ref{fig:energy_ev} shows the evolution of $e_{\rm MRI}$ and $e_{\rm PI}$ for different values of $C$ using the MInIT model. While at saturation $e_{\rm MRI}$ and also $e_{\rm PI}$ do depend on the value of $C$, there is no such dependence during the growth phase. 

 Our goal is to find the optimal value of $C$ that minimizes the differences between the results of the direct numerical simulation and our model. To estimate these differences we use the $L_2$ norm (relative error) defined as
\begin{equation}\label{l2norm}
    L_2 = \sqrt{\frac{1}{2}\sum_i(x^i_{\rm s}-x^i_{\rm m})^2\Bigg(\frac{1}{\sum_j{(x^j_{\rm s})^2}}+\frac{1}{\sum_j{(x^j_{\rm m})^2}}\Bigg)}\,,
\end{equation}
where $x^i_{\rm s}$ are the data from the numerical simulation, and $x^i_{\rm m}$ are the data from the model. To obtain a relative error in Eq.~\eqref{l2norm} we normalize its value to the harmonic mean between the average value of $x^i_{\rm s}$ and $x^i_{\rm m}$. Expressed in this way, $L_2$ will be large when $x_{\rm s}$ and $x_{\rm m}$ differ considerably (several orders of magnitude) and will be smaller than unity when $x^i_{\rm s} = x^i_{\rm m}+\delta $, with $\delta \ll 1$. Having values with the same order of magnitude will give $L_2\sim 1$. The optimal value of $C$ is obtained by minimizing the relative error from Eq.~\eqref{l2norm} with $C$ as a free parameter, using the stress tensors as our data. The $x^i_{\rm m}$ set is composed by the stress tensors computed with the $\alpha$, $\beta$ and $\gamma$ coefficients and the evolved stress energy densities, $e_{\rm MRI}(t)$ and $e_{\rm PI}(t)$. On the other hand, the $x^i_{\rm s}$ set consists of the stress tensors directly obtained by the filtering of the output of a numerical simulation and the application of Eq.~\eqref{stress_tensors}. Moreover, we only consider the saturation part of the simulation ($t>12$~ms), which is where parameter $C$ plays an important role (see Fig.~\ref{fig:energy_ev}). 

In order to have a proper statistical error in the computation of the $L_2$ norm, we need an ensemble of points on which to compute the norm. 
A way to obtain reliable results with this procedure is to apply the box filter of Eq.~\eqref{box_filter} over a reasonably large number of grid cells, in our case $10\times 10\times 10$ cells located in the center of the box. The filter size must be large enough so that $\Delta_f > \lambda_{\rm MRI}$, where $\lambda_{\rm MRI}$ is the wavelength of the fastest growing mode of the instability. Using a filter smaller than the fastest growing mode would capture different modes which would lead to different values for the growth rates. From Table \ref{tab:table_sim} we see that the size of the box is at least 3 times larger than $\lambda_{\rm MRI}$ in the $r$ and $z$ directions and 12 times larger in the angular direction. Depending on the resolution, $\lambda_{\rm MRI}$ will be a certain number of times larger than the size of the computational cell, $\Delta$. In the highest resolution simulation, MRI-H1, $\lambda_{\rm MRI} = 33.3 \Delta$. Thus, a filter of size $S_f = \Delta_f/\Delta = 40$ leads to $\Delta_f = 1.2 \lambda_{\rm MRI}$. For completeness, we also apply two more box filters with sizes $S_f = 50$ and $S_f = 60$. We do this for all resolutions since all these filter sizes satisfy $\Delta_f > \lambda_{\rm MRI}$. In addition, for the simulations MRI-L1 and MRI-M1 we also use $S_f = 30$. However, for the simulation MRI-L1, $S_f = 60$ yields a filter equal to the size of the box and, thus, this filter is not used for that simulation. Given this, the scales in which the parasitic instabilities are developed will be represented by $\lambda_{\rm MRI}$ since $\Delta_f > \lambda_{\rm MRI}$, and Eq.~\eqref{std} will have $\lambda = \lambda_{\rm MRI}$. 

\begin{table}
  \centering
  \begin{tabular}{|c|c c c|}
\hline
    \multicolumn{4}{c}{\textbf{Optimal values of the $C$ parameter}} \\
    \hline
    \hline
       &  Maxwell & Reynolds & Mean  \\
    \hline
    $S_f = 40$ & $8.2^{+ 0.9}_{ - 0.9} $ & $8.5^{+ 0.6} _{- 0.9}$ & $8.3^{+0.7}_{ -0.9}$   \\
    \hline
   $S_f = 50$ & $8.5 ^{+ 0.9}_{ - 0.6}$ & $8.8^{ + 0.6}_{ - 0.9}$ & $8.6^{+ 0.7}_{ -0.7}$    \\
    \hline
    $S_f = 60$ & $8.8^{ +0.9}_{ - 0.6}$ & $8.8^{+ 0.9}_{ - 0.9}$ & $8.8 ^{+ 0.9}_{ - 0.7}$  \\
    \hline
\end{tabular}
\caption{Optimal values of the parameter $C$ that minimize the $L_2$-norm of the Maxwell and Reynolds stress tensors. The last column reports the mean values of the two, for different filter sizes.}
\label{tab:table_factor_sat}
\end{table}

\begin{figure*}
    \centering
    \includegraphics[width = \textwidth]{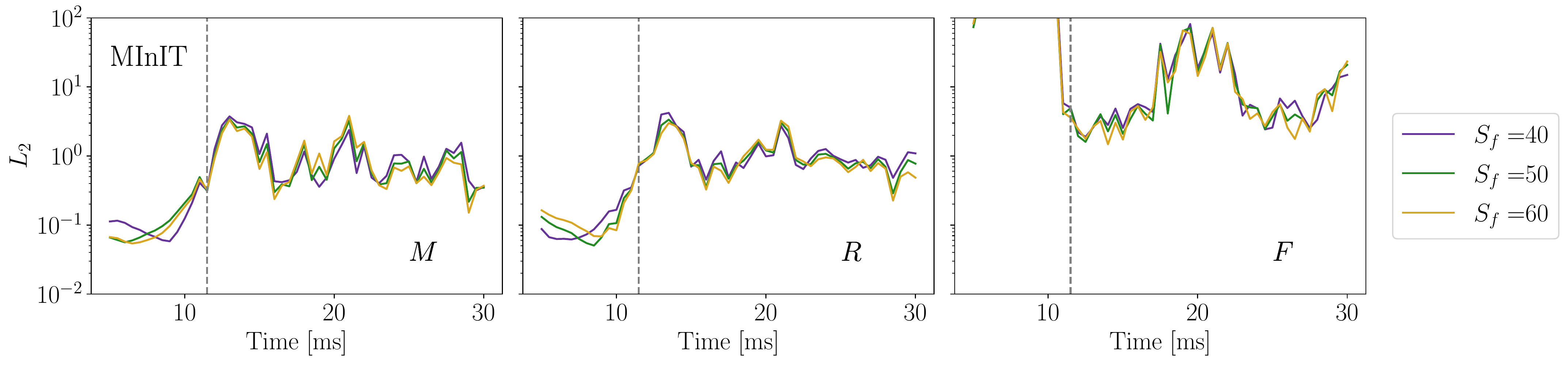}
    \includegraphics[width = \textwidth]{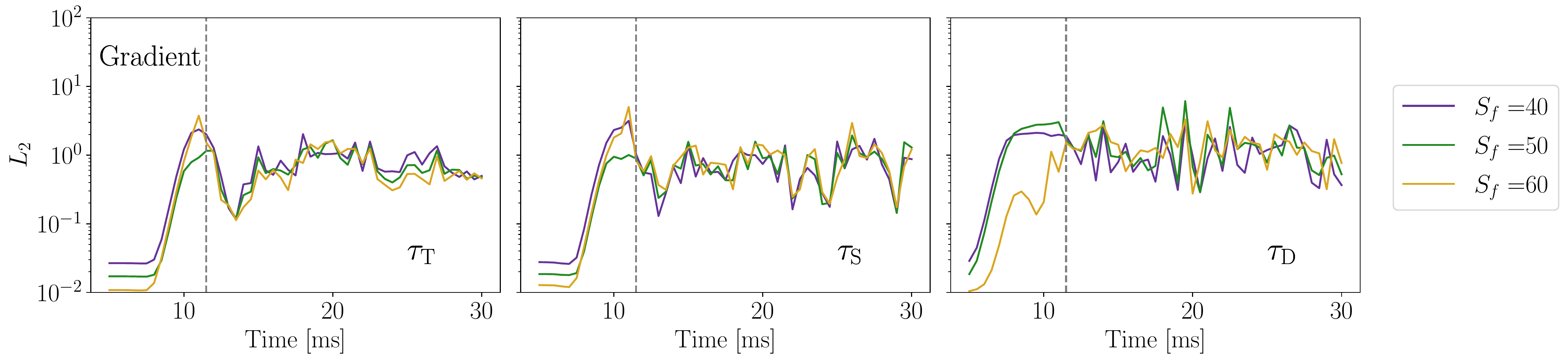}
    \caption{Time evolution of the $L_2$-norm for both the MInIT model (top panels) and the gradient model (bottom panels), using simulation MRI-H1. The vertical dashed lines signal the saturation time of the instability. For the MInIT model, the $L_2$-norm is below $\sim 5$ for the Maxwell and Reynolds stress tensors while for the Faraday stress is around 10 after saturation. Similar values are found for the SFS tensors of the gradient model for all cases.}
    \label{fig:l2_time_ev}
\end{figure*}

We calculate the $L_2$-norm by applying Eq.~\eqref{l2norm} to the modelled and simulation-based stresses. More specifically, we obtain the $L_2$-norm for each time iteration by making the summation in Eq.~\eqref{l2norm} over the spatial points and all the components of the stresses, in order to give more weight to the larger components, and then we compute the root mean square of the result over time. Table~\ref{tab:table_factor_sat} reports the values of $C$ that minimize the $L_2$-norm for the Maxwell and Reynolds stress tensors\footnote{We do not use the Faraday stress tensor because all its components are much smaller than the ones from the other stresses.} for different filter sizes, using the highest-resolution simulation, MRI-H1. The upper and lower bounds indicate 10$\%$ variations in the minimized $L_2$-norm. The last column reports the mean value of $C$. It slightly grows with the size of the filter but all values fit inside the different confidence intervals. Averaging over the filter sizes, we obtain
\begin{equation}\label{c_optimal}
    C_{\rm opt} = 8.6 \pm 0.8 \,.
\end{equation}

The same constant is used even when applying the model to other simulations with different resolution.

\subsection{Test of the sub-grid models: an a-priori test}

\subsubsection{Preliminaries}

After calibrating the coefficients of the MInIT model we turn next to assess the performance of our new sub-grid model compared to the gradient model. This will be done through a so-called a-priori test. This consists in applying a filtering operation (see Section~\ref{sec:filter}) to data from a numerical simulation and compute from these data the corresponding terms of the sub-grid model to test. This allows for a quantitative comparison with the terms computed by the analytical model, e.g.~using Eqs.~\eqref{sfs_tensors} for the gradient model or Eqs.~\eqref{closure_coeffs} for our sub-grid model. 

A way to check the goodness of a model is to compute the $L_2$-norm between the data from the simulation and that from the model (see Eq.~\eqref{l2norm}). We note that~\cite{paper_gradient_test} used the so-called Pearson correlation coefficient instead of the $L_2$-norm as the metric to assess the quality of the gradient sub-grid model. This coefficient, however, turned out not to be suitable for the assessment of our model. While the Pearson coefficient measures the linear correlation between two sets of data, it is not necessary for a model to have a strong linear correlation in order to fit well to data. The goal of the MInIT sub-grid model is not to have tight correlations in the evolution of the different quantities but to provide a representation that is statistically representative of the different quantities on average. For this reason, we 
resort here to the $L_2$-norm metric for the model assessment. Nevertheless, 
in Appendix~\ref{sec:appendix} we report the results obtained for the gradient model using the Pearson correlation coefficient as well. This allows for a comparison with the results reported by~\citet{paper_gradient_test} for the KHI and see whether they are consistent with our findings for the case of the MRI.

\subsubsection{A-priori test of the models}

In the a-priori test we compute the $L_2$-norm of the difference between the numerical data, $x^i_{\rm s}$, and the data obtained with the evolution equations of the model, $x^i_{\rm m}$. This is similar to what we did in Section~\ref{sec:OPTC} but now we make use of the whole simulation, i.e.~considering the growth phase of the instability as well. 
As before, we only apply filters with size $\Delta_f > \lambda_{\rm MRI}$ since we are considering the fastest growing mode of the instability. Filter sizes $S_f = 40$, 50 and 60 will be also employed in the test of the gradient model to do a comparison between both models. 

Fig.~\ref{fig:l2_time_ev} shows the  values of the $L_2$-norm for both models before the time-average is performed. The top row  corresponds to the MInIT sub-grid model while the bottom row depicts the results of the gradient model. In all the plots in the figure we ignore the first 2 ms of the simulations in order to get rid of initial transients. As the figure shows, the values of the $L_2$-norm are below $\sim 5$ at most times and for the two sub-grid models, except for the Faraday stress tensor from the MInIT model which reaches values larger than $10^2$ during the initial growth phase. For both models the largest values of the $L_2$-norm are attained during the saturation phase, as expected. For the MInIT model the coefficients $\alpha^{\rm MRI}$, $\beta^{\rm MRI}$ and $\gamma^{\rm MRI}$ are analytical and match almost perfectly the simulation-based stresses during the exponential growth. In this phase they dominate because $e_{\rm MRI} \gg e_{\rm PI} $. At saturation $e_{\rm PI} \sim e_{\rm MRI}$, and thus the calibrated coefficients $\alpha^{\rm PI}$, $\beta^{\rm PI}$ and $\gamma^{\rm PI}$ also play a role.  

\begin{figure}
    \centering
    \includegraphics[width = \linewidth]{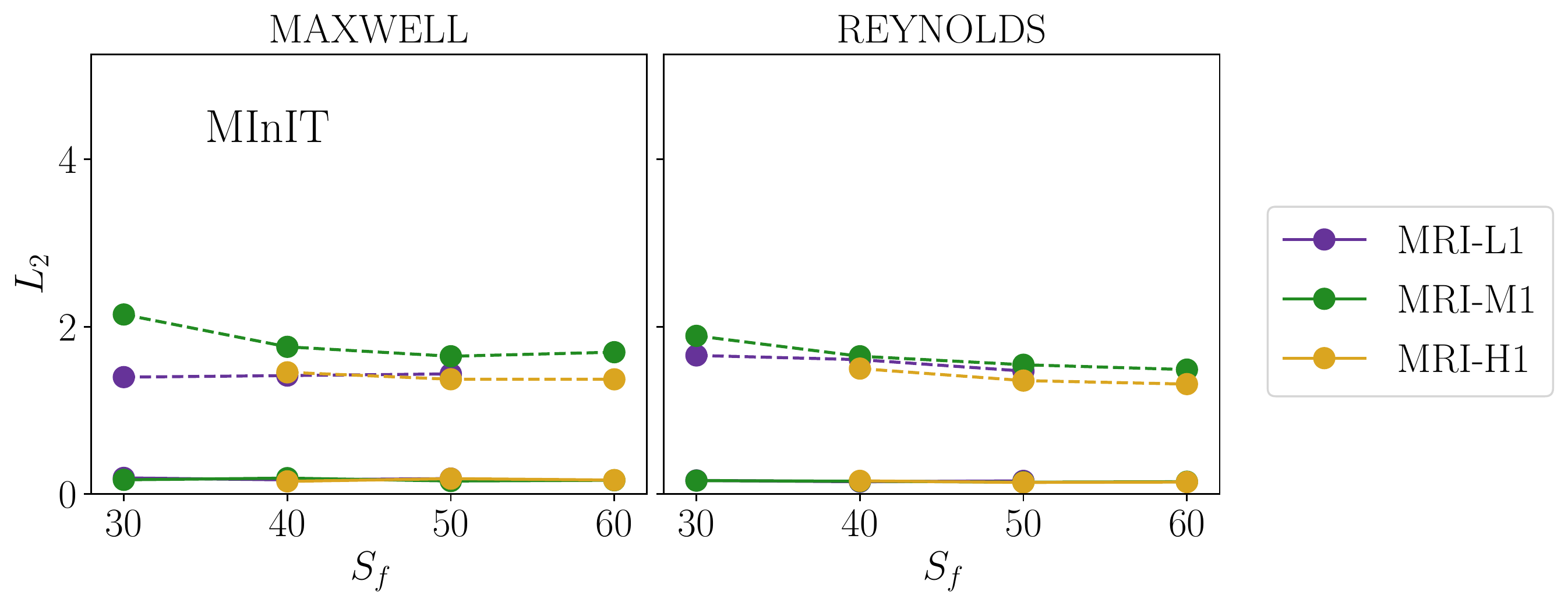}
    \includegraphics[width = \linewidth]{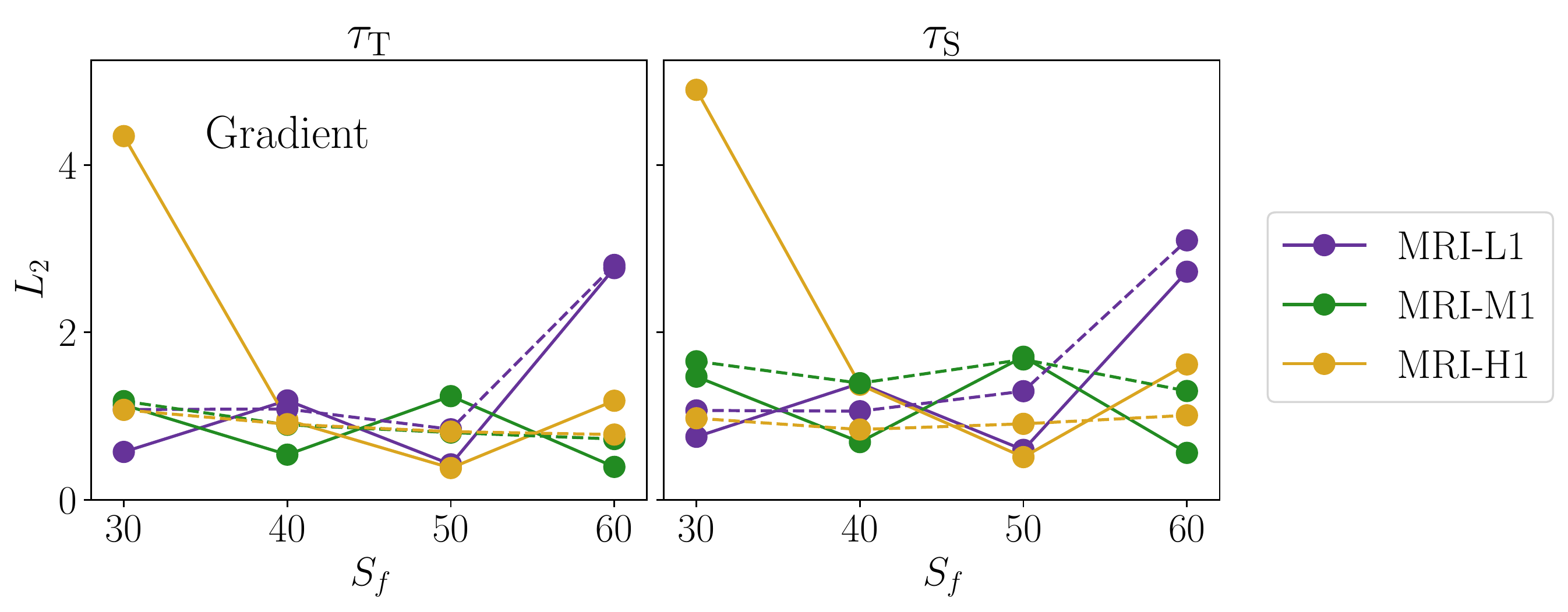}
    \caption{$L_2$-norm of the Maxwell and Reynolds stress tensors of the MInIT model (top row) and of the $\tau_{\rm T}$ and $\tau_{\rm S}$ SFS tensors of the gradient model (bottom row) for different filter sizes, computed over space and time-averaged. Solid (dashed) lines correspond to the initial exponential growth (saturation) of the instability. Colours correspond to simulations with different resolutions, as indicated in the legend.}
    \label{fig:l2_sf_MR}
\end{figure}

\begin{figure}
    \centering
    \includegraphics[width = \linewidth]{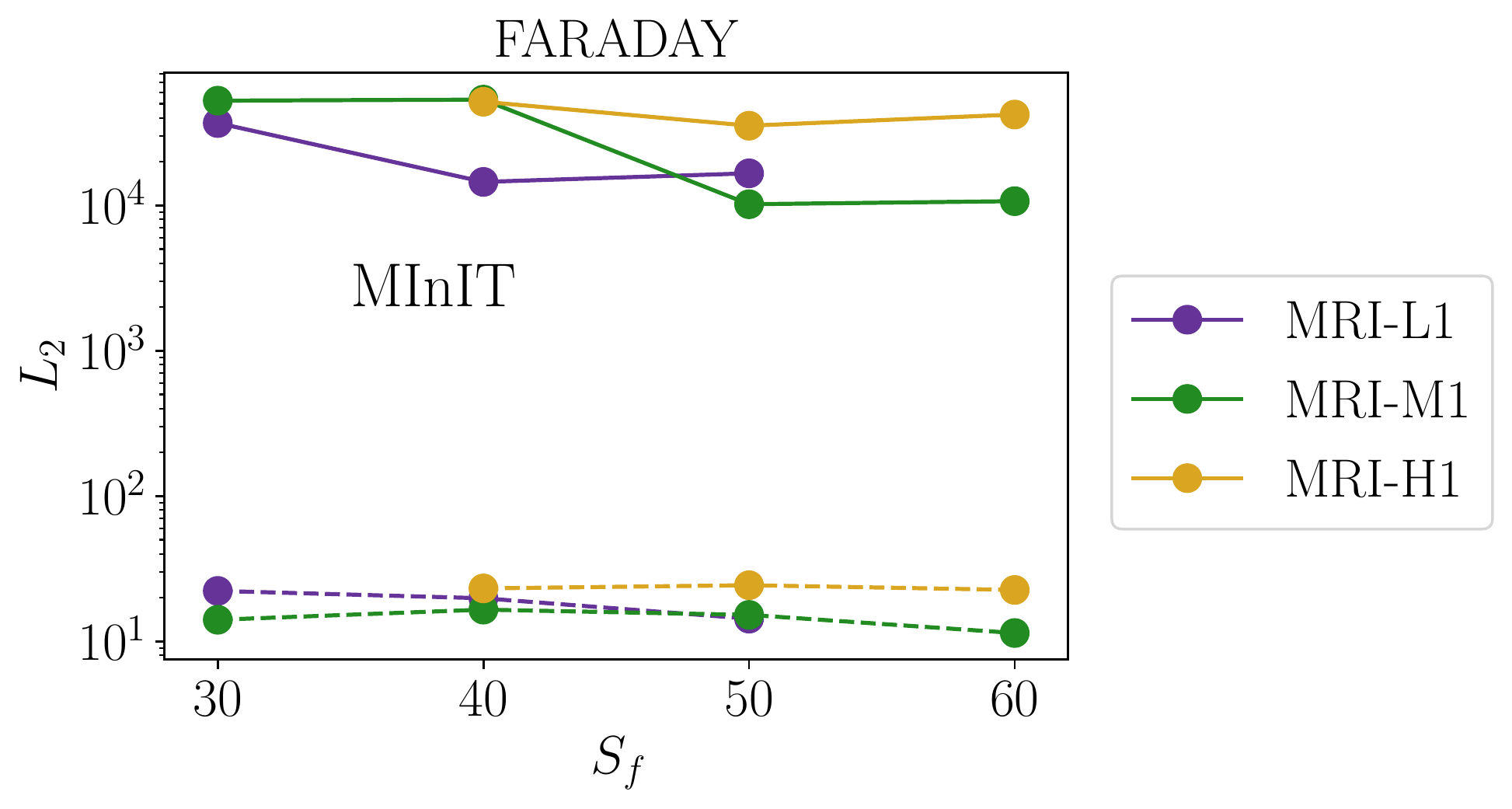}
    \includegraphics[width = \linewidth]{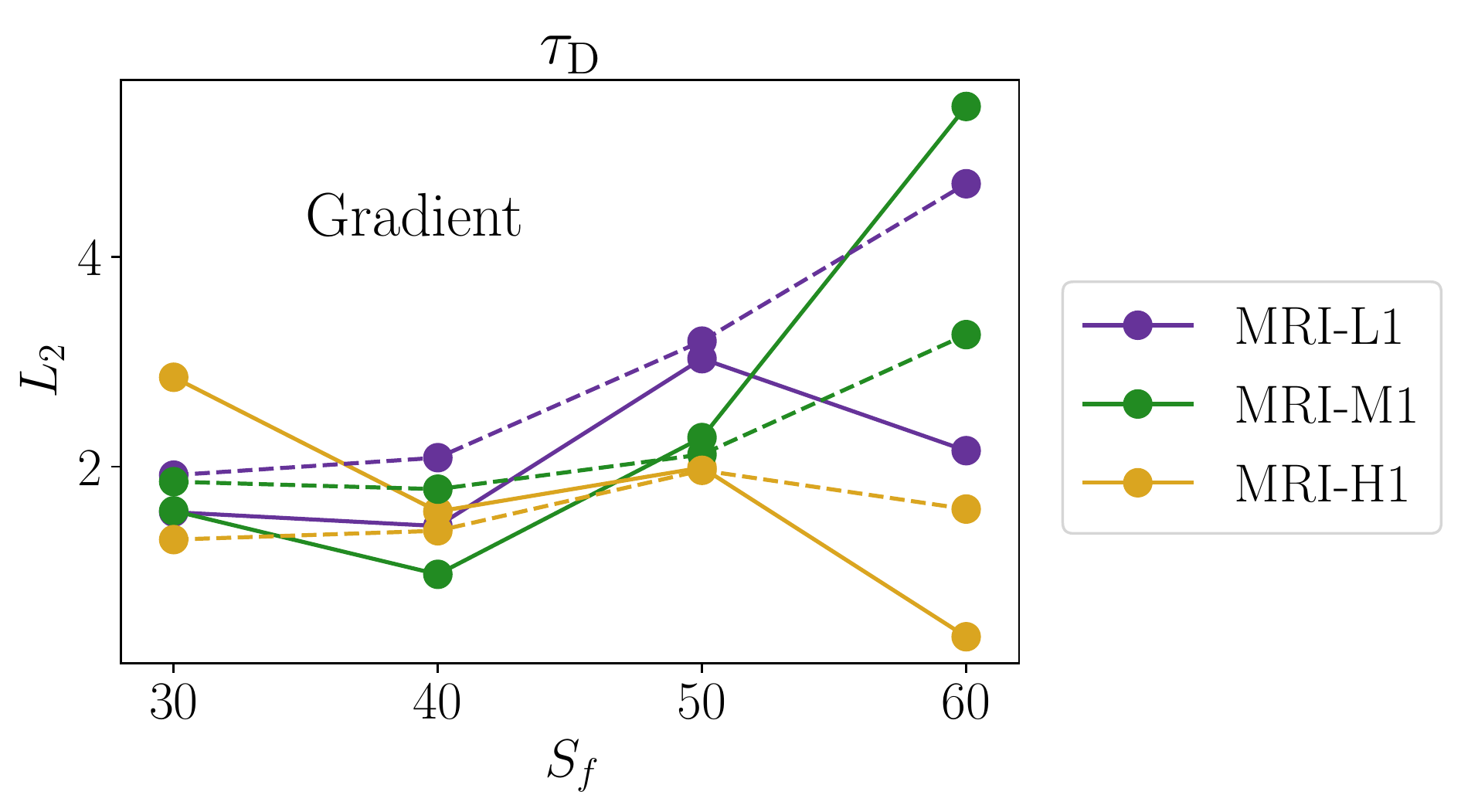}
    \caption{As Fig.~\ref{fig:l2_sf_MR} but for the Faraday stress tensor of the MInIT model (top row) and the $\tau_{\rm D}$ SFS tensor of the gradient model (bottom row).}
    \label{fig:l2_sf_F}
\end{figure}

In Figs.~\ref{fig:l2_sf_MR} and \ref{fig:l2_sf_F} we depict the $L_2$-norm computed over space and averaged in time by means of the root-mean-square for each filter size. Solid lines correspond to the initial exponential growth of the instability and dashed lines to the saturation phase. As in Fig.~\ref{fig:l2_time_ev}, the top row of both figures depicts the values obtained for the MInIT model and the bottom row those of the gradient model. For the quantities reported in Fig.~\ref{fig:l2_sf_MR} (Maxwell and Reynolds stress tensors and $\tau_{\rm T}$ and $\tau_{\rm S}$ SFS tensors) we find that the values of the $L_2$-norm are ${\cal O}(1)$ for both sub-grid models. This indicates that the two models fit well the data of the simulations, i.e.~model and data differ by less than an order of magnitude from each other. In Fig.~\ref{fig:l2_sf_F} we show the special case of the Faraday stress tensor and its analogue tensor ($\tau_{\rm D}$) computed with the gradient model. The $L_2$-norm of the Faraday tensor reaches values larger than $10^4$ during the growth phase due to the artefacts already discussed in Section~\ref{sec:PI}, since the components of this tensor are exclusively given by $e_{\rm PI}$. After saturation, the $L_2$-norm of the Faraday stress tensor decreases to values of $\mathcal{O}(10)$. Correspondingly, the values attained by the $\tau_{\rm D}$ SFS tensor of the gradient model are of $\mathcal{O}(1)$.

Figs.~\ref{fig:l2_sf_MR} and \ref{fig:l2_sf_F} exhibit that the MInIT model shows almost no dependence on the filter size, for all stress tensors. If anything, the norm is even slightly smaller for larger filters. Moreover, simulations with different resolutions yield almost the same values of the norm~\footnote{Simulations with even finer resolutions than the ones employed in this work would be needed to determine whether there could be an actual dependence on resolution.}. On the other hand, for the gradient model the behaviour of the norm with the filter size is markedly different. In most cases, in particular for low-resolution simulations, the norm of the SFS tensors increases with filter size. This is consistent with the results reported by~\cite{paper_gradient_test} who, for the case of the KHI, found a similar behaviour for the  Pearson coefficient for different filter sizes (a similar study with this coefficient for the MRI is shown in the Appendix~\ref{sec:appendix}). 

\section{Discussion and conclusions}
\label{sect: sec5}

Time-dependent, direct numerical simulations of astrophysical systems (and in other fields too) have limitations to capture the dynamics at all scales of interest. In particular, the correct description of the development of turbulence at small scales is a challenge for current grid-based simulations which typically suffer from insufficient resolution. Instabilities such as the magnetorotational instability and the Kelvin-Helmholtz instability play a major role in the amplification of weak magnetic fields of the post-merger remnant of binary neutron star mergers. Its correct modelling is paramount for reliable estimates of the lifetime of a hypermassive neutron star and of the multi-messenger observational signatures thereof. Linking the results of simulations with the wealth of new data from multi-messenger observations of BNS mergers, sGRBs, and kilonovae, is still an ongoing task.

Despite continuous progress in the computational front, with ever more efficient and accurate numerical methods and treatment of physical processes, it is still not feasible to reproduce all physical effects involved in certain astrophysical scenarios through direct numerical simulations. An alternative to these computationally expensive simulations are the so-called sub-grid models, which try to deal with the effects of the small scales in terms of the resolved scales, with modest resolution. In this paper we have assessed different sub-grid models using three-dimensional box simulations of the MRI~\citep{Rembiasz-2016}.

The first model we have tested is the $\alpha,\beta$-dynamo model. This is a fairly simple (and limited) model that applies several assumptions to reproduce the dynamo effect arising in the induction equation for the magnetic field. We have found that once the flow is fully turbulent, the different components of the $\alpha$-dynamo reach a similar constant value. However, setting a constant value for this coefficient throughout the development of turbulence would not work as it would not capture properly the exponential growth of the instability.

The second sub-grid model we have tested is the gradient model. Some recent studies of BNS mergers have implemented this model \citep[e.g.][]{2021arXiv211208413P} with promising outcomes - the turbulent amplification of the magnetic field obtained with the model is similar to that obtained with direct numerical simulations using twice the resolution. The gradient sub-grid model is simulation agnostic as no physical or phenomenological assumption is made since its closure relation to model the sub-filter-scale tensors is based on the Taylor expansion and the inverse function theorem. Thus, this model is universal for any kind of astrophysical scenario and it is not limited by the physical properties of the problem. 

The focus of this paper has been to present and assess a 
new sub-grid model, the MHD-instability-induced-turbulence (MInIT) mean-field model. The main appeal of the MInIT model is that it is physically motivated as it is based on the time evolution of the turbulent stress tensors and their relationship  with the turbulent energy density of the MRI and of the parasitic instabilities. By considering a simple linear dependence between the tensors and the energy densities, the model only needs two partial-differential evolution equations for the energies to compute all quantities. These equations take into account the effect of the parasitic instabilities that saturate the growing turbulence, and also their dissipation at the end of the Kolmogorov cascade, which makes the turbulent stresses to exponentially grow up to a saturation value. The equations also take into account the growth rate of the (fastest growing mode of the) MRI and also of the PI as functions of resolved quantities.

Once the evolution equations are solved, the turbulent stress tensors are obtained by using the constant coefficients that link them to the energy densities. Those are obtained from control numerical simulations and they are found to be almost equal for the range of resolutions and initial magnetic fields considered in this work. While these coefficients seem therefore universal (at least for simulations of the same type of instability, but see the discussion below) due to the isotropy and homogeneity that arise from the turbulent dynamics, further studies with different initial configurations may be needed to confirm this. We note that, contrary to the gradient sub-grid model, in our new model no spatial derivatives need to be computed and most quantities used in the evolution equation are global parameters of the simulation that we have control on.

The MInIT sub-grid model has been assessed through an a-priori test, i.e.~using data from a direct numerical simulation and applying a filtering operation to compare the filtered data with that given by the model. We have used the $L_2$-norm (relative error) as our metric to quantify the comparison, obtaining values below $\sim 5$ for most cases. This means that the data from the simulation and the modeled stresses do not differ more than one order of magnitude. An order-of-magnitude agreement is certainly an achievement for such a simple model and should be sufficient for its use in complex numerical simulations where even larger uncertainties arise from the modelling of many of the physical ingredients (e.g. equation of state or neutrino transport). Since the Faraday tensor is in average compatible with zero, the large relative errors observed in the Faraday tensor for the MInIT model should in principle not be a problem for the applicability of the model to global simulations, where its effect on the dynamics would be small. However, given the possible role of the Faraday tensor in the formation of large-scale dynamos, future studies and extensions of the MInIT model could focus in a better description of this component.

Moreover, no dependence on the filter size or the length scale of the unresolved scales has been found, as opposed to the gradient model in which the $L_2$-norm (slightly) increases with the filter size, particularly for low-resolution simulations. For an ideal sub-grid model, there should not be a dependence on the filter size or on the typical length of sub-grid scales, and it should also properly work in the limit $S_f \rightarrow \infty$. This limit represents the case in which the filter is applied to a fully resolved simulations, i.e. with "infinite" resolution. We have also observed that the MInIT model behaves consistently for simulations with different resolutions and initial magnetic fields, as those have yielded similar values of the $L_2$-norm of the stress tensors. 

In its comparison with the gradient model, the MInIT model seems to perform with comparable accuracy (except for the Faraday tensor discussed above). This comparison is however unfair and, despite of the appearances, different things are being tested and compared. For the case of the MInIT model, the only required information from the simulation is the initial value of the mean quantities, and from those, the rest of the evolution of predicted. This process mimics the case of its application to global simulations in whose the only information known is the average values at grid cells and all the dynamics at sub-grid scales should be modelled. In numerical simulations of the MRI, if the grid cell size is not sufficiently small the instability will not be captured and turbulence will not develop (or will do it at a slower rate). The MInIT model, allows to model MRI in this sub-grid scales and the development of turbulent stresses even if resolution is not sufficient. In opposition, the test for the gradient model uses the time evolution of the average quantities over the filter size (not only the initial values). Since the MRI is well resolved in the simulation the mean quantities evolve in time and the prediction of the gradient model follows this evolution. If only the initial values were provided to the gradient model, the model would catastrophycally fail to predict the growth of the MRI. Somehow, the gradient model needs that the MRI is minimally resolved, and only then is capable of describing the turbulence at even lower scales. Therefore, a better statement for the comparison between both model is that the MInIT model is capable of achieving similar results than the gradient model with less information from the simulation. This may imply that in global simulations the MInIT model may need a lower resolution to achieve the same results that the gradient model. However, this should be tested in the future.

The theoretical framework developed in this work can be applied to different astrophysical systems. Resolving the MRI is a subject of interest across different areas of astrophysics, specially in the modelling of magnetized discs at all scales, e.g.~discs around compact objects, protostellar and proto-planetary discs, and those systems could benefit of this model. However, extrapolating the calibrated coefficients obtained in this work to other situations should be handled with care. In particular our model assumes a regime in which the Reynolds number is large, the  magnetic field is not dominant, and turbulence is approximately incompressible. If these conditions are fulfilled, our theoretical arguments suggest that the coefficients could be used outside the range explored here, although proper testing and re-calibration would be encouraged. If some of those conditions do not hold, the coefficients may have different values and/or additional dependencies. Our most immediate aim is to assess the possible universality of the model when applied to a different kind of instability (e.g.~the Kelvin-Helmholtz instability) as well as to further improve the  model by a deeper investigation of the relationship between the stress tensors and the turbulent energy densities.

Applications envisaged include the study of  magnetic-field amplification in proto-neutron stars following a core-collapse supernovae and in HMNS resulting from BNS mergers. Sub-grid models as the one reported here can greatly help direct numerical simulations by properly capturing the magnetic-field amplification at small scales which has potential implications on the lifetime and dynamics of highly-magnetized astrophysical compact objects. For this purpose, our MInIT model should be seen as a complex and time-dependent sub-grid model that acts as a closure model. Using this time-dependent closure in combination with an augmented system of MHD equations, similar to the one proposed by \cite{ogilvie}, global simulations should be possible in a similar fashion to large-eddy simulations.

\section*{Acknowledgements}

Work supported by the Spanish Agencia Estatal de Investigaci\'on (Grants No. PGC2018-095984-B-I00 and PID2021-125485NB-C21) and by the Generalitat Valenciana (Grant No. PROMETEO/2019/071). MMT acknowledges support by the Spanish Ministry of Universities through the FPU Ph.D.~grant No.~FPU19/01750.
MO acknowledges support  from the Spanish Ministry of Science, Innovation and Universities via the Ram\'on y Cajal programme (RYC2018-024938-I).

\section*{Data Availability}
The data underlying this article will be shared on reasonable request to the corresponding author.


\bibliographystyle{mnras}
\bibliography{draft} 



\appendix

\section{Pearson correlation coefficient for the gradient model}
\label{sec:appendix}
\begin{figure*}
    \centering
    \includegraphics[width=0.7\textwidth]{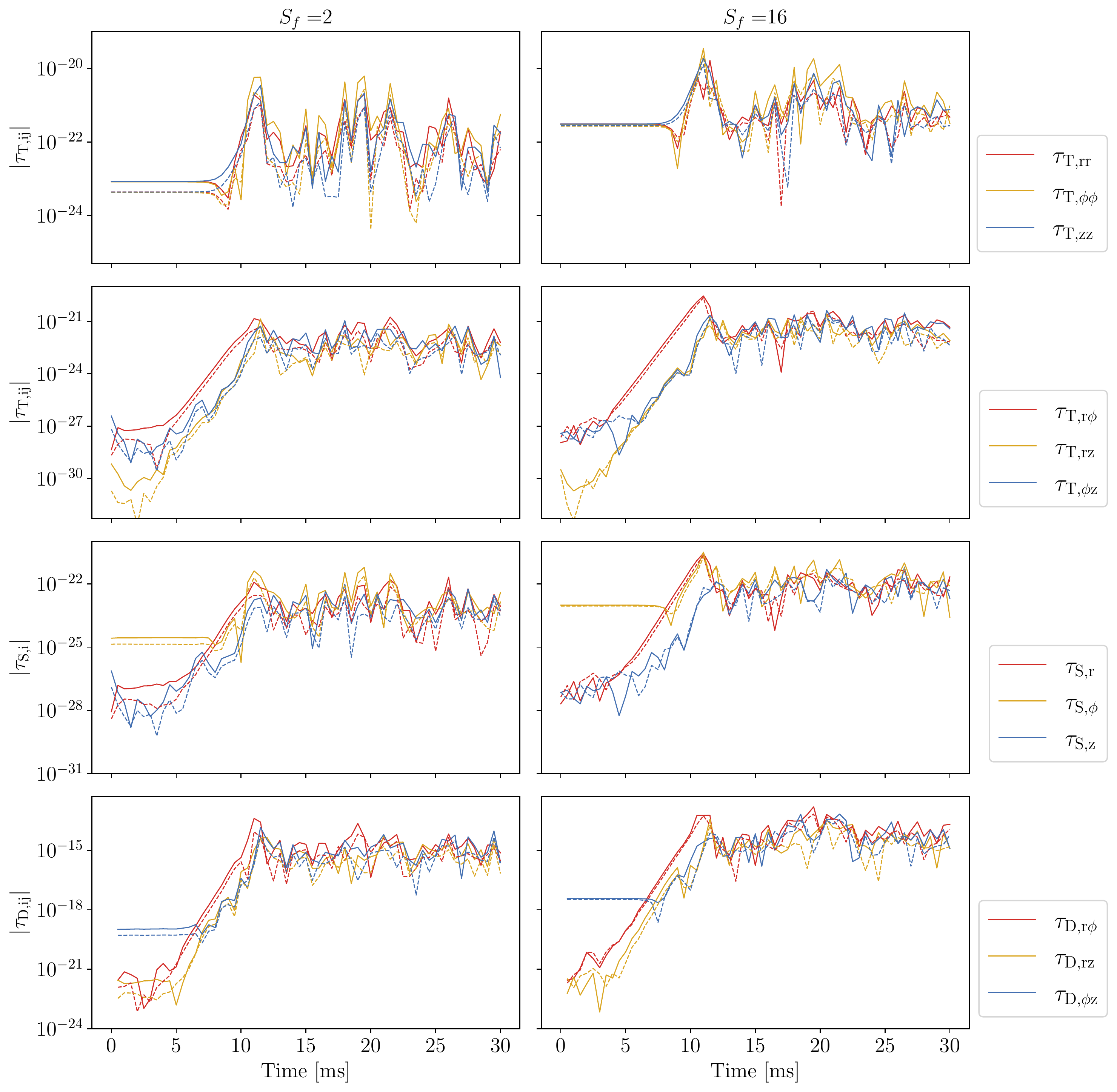}
    \caption{Comparison of the sub-filter-scale tensors from simulation MRI-H1 using different filter sizes, $S_f = 2$ (left column) and $S_f = 16$ (right column). The solid lines represent the simulation-based values and the dashed line, the modeled quantities.}
    \label{fig:sfs_tensors}
\end{figure*}

We discuss in this Appendix the results of the a-priori test for the gradient model using as metric the Pearson correlation coefficient, to compare with
 \cite{paper_gradient_test}. These authors used data from simulations of the KHI. Here we employ our own MRI box simulations. The Pearson coefficient measures the linear correlation between two sets of data, $x_i$ and $y_i$, of size $N$. It is given by
\begin{equation}\label{pearson_definition}
    r_{x,y} = \frac{N\sum_i x_i y_i -\sum_ i x_i \sum_i y_i }{\sqrt{N\sum_i x_i^2-(\sum_i y_i)^2}\sqrt{N\sum_i y_i^2-(\sum_i y_i)^2}}\, , 
\end{equation}
and it takes values from $-1$ (complete negative correlation) to $+1$ (complete positive correlation). We consider that the model fits well the data if the correlation coefficient is $+0.6$ or larger. The data sets we use to compute the coefficient are the spatial values of the SFS tensors. We average over the different independent components of each tensor, since the differences between components are statistically negligible \citep{extension_subgrid_vigano,paper_gradient_test}. The time evolution of the SFS tensors for two different filters is plotted in Fig.~\ref{fig:sfs_tensors}. We next average in time from the instant in which turbulence is completely developed, leading to a stationary regime.

\begin{figure}
    \centering
    \includegraphics[width=0.9\linewidth]{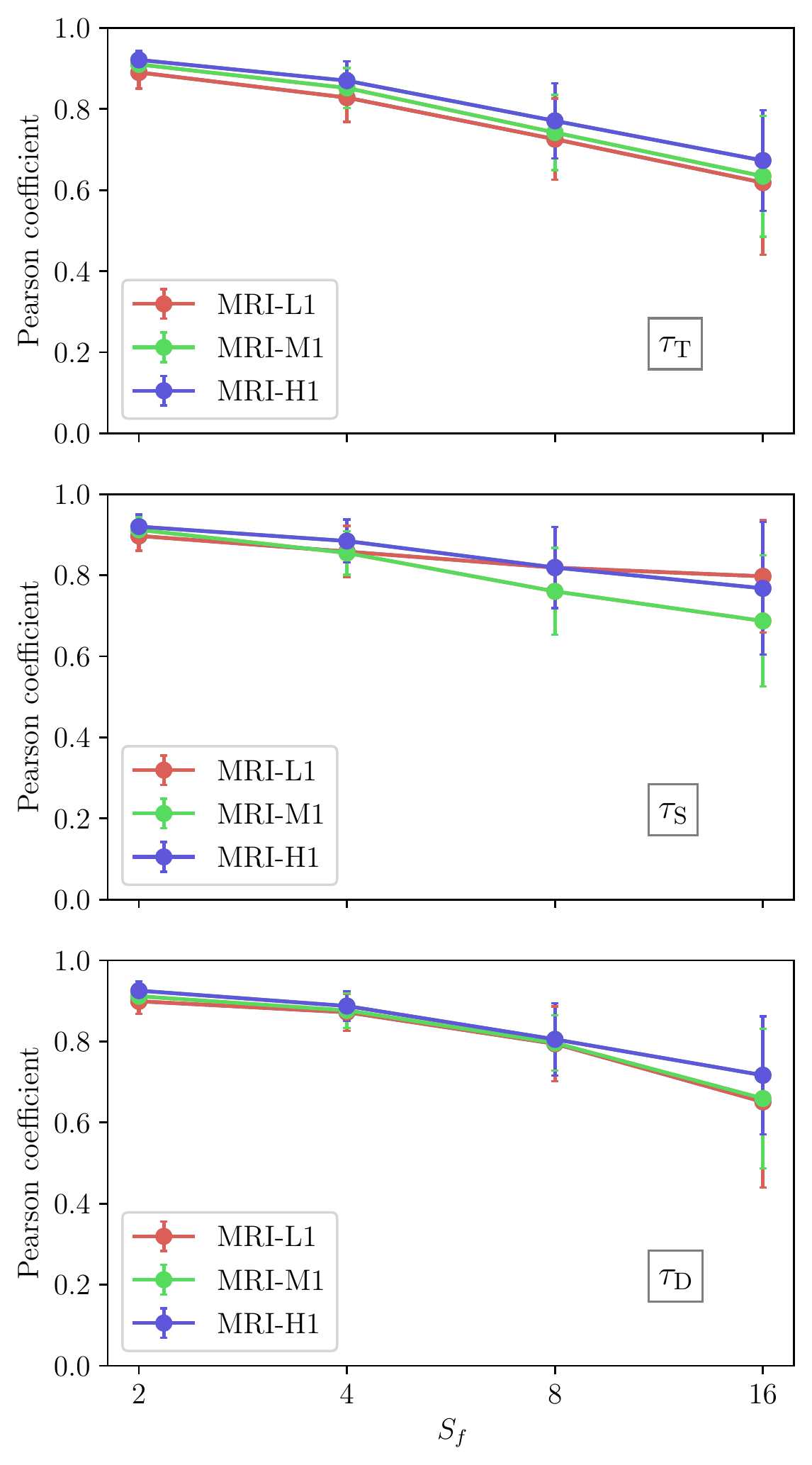}
    \caption{Values of the Pearson correlation coefficient for each filter size, $S_f$, and for each SFS tensor. The correlation  is computed over space in the saturation regime and averaging in time.}
    \label{fig:sf_comparison}
\end{figure}

\begin{figure}
    \centering
    \includegraphics[width=0.9\linewidth]{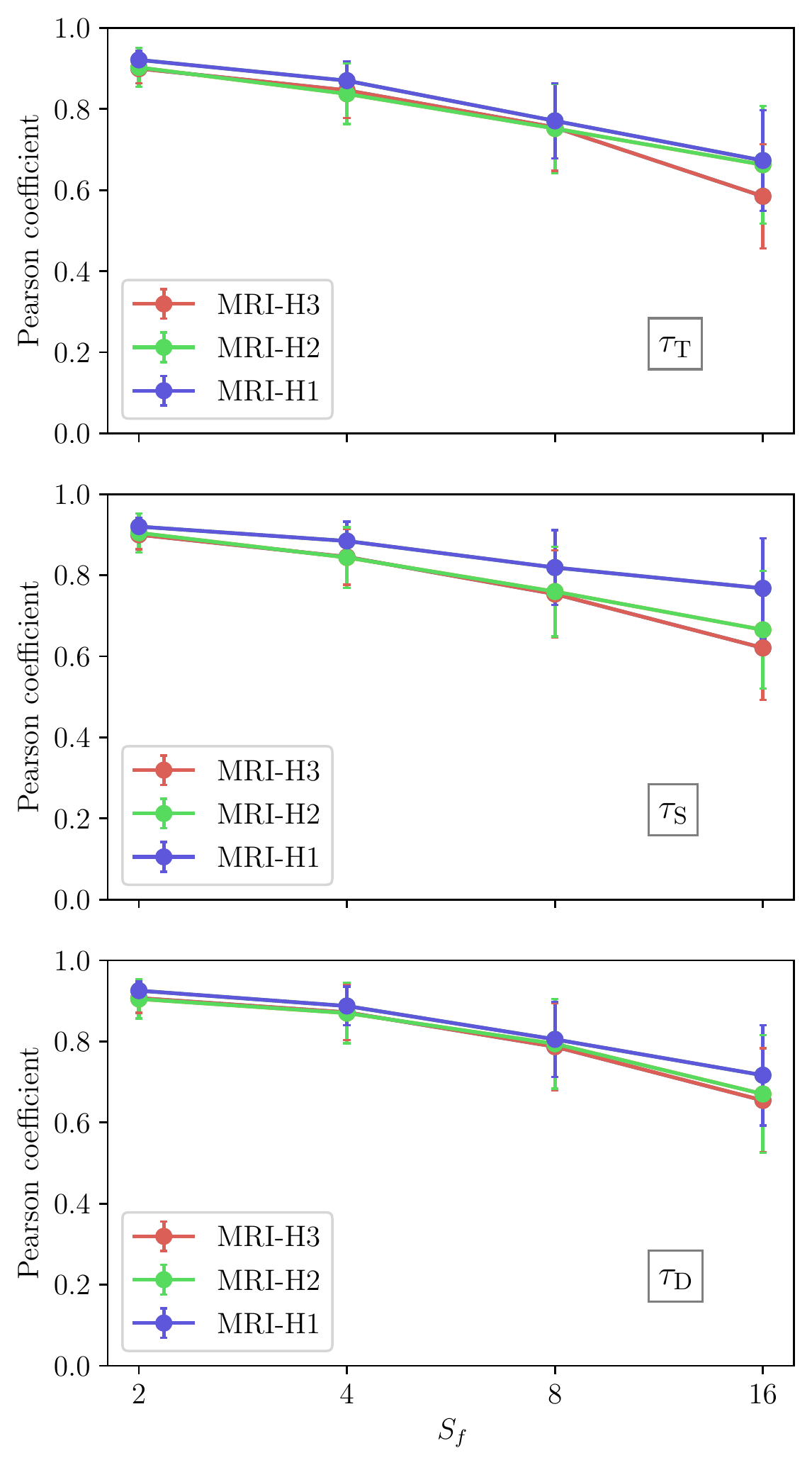}
    \caption{Values of the Pearson correlation coefficient for each filter size, comparing simulations with different initial magnetic field strengths and keeping the same numerical resolution. The correlation is computed over both space and time, as in Fig.~\ref{fig:sf_comparison}.}
    \label{fig:sf_comp_B}
\end{figure}

The values of the Pearson correlation coefficient are plotted in Figs.~\ref{fig:sf_comparison} and \ref{fig:sf_comp_B}. On the one hand, 
Fig.~\ref{fig:sf_comparison} shows that, for all SFS tensors, the coefficient is larger for smaller filter sizes, with correlation values close to 0.9. As the filter becomes larger, the correlation degrades to values close to 0.6. Moreover, as the resolution of the numerical simulation increases, so does the Pearson correlation coefficient. 
On the other hand, Fig.~\ref{fig:sf_comp_B} shows that the correlation hardly changes for different initial magnetic field amplitudes. It is only slightly larger for simulation MRI-H1, the one with the strongest initial magnetic field of our sample. Moreover, for all simulations the correlation coefficient decreases as the filter size increases. 

The results from this Appendix show that the trends we have found for the Pearson correlation coefficient using box MRI simulations are in agreement with the results reported by \cite{extension_subgrid_vigano} and~\cite{paper_gradient_test} using box KHI  simulations.

\bsp	
\label{lastpage}
\end{document}